\documentclass[a4paper, 12 pt]{article}\usepackage{a4wide}
\usepackage{amsmath}
\usepackage{amsfonts}
\usepackage{amsthm}
\usepackage{amscd} 

\numberwithin{equation}{section}

\newcommand{\half}{{\textstyle \frac{1}{2}}}

\newcommand{\fhalf}{{\textstyle \frac{5}{2}}}

\newcommand{\ui}{\textrm{i}}
\newcommand{\UI}{\textrm{I}}

\newcommand{\ue}{\textrm{e}}

\newcommand\ttimes{{\times}}% \ttimes is just \times, but is used to make 
\newcommand\alphavec{{\boldsymbol \alpha}}
\newcommand\betavec{{\boldsymbol \beta}}
\newcommand\gammavec{{\boldsymbol \gamma}}

\newcommand\muvec{{\boldsymbol \mu}}
\newcommand\sigmahat{{\hat \sigma}}
\newcommand\uhat{{\hat u}}
\newcommand\thetahat{{\hat \theta}}
\newcommand\xihat{{\hat \xi}}
\newcommand\psihat{{\hat \psi}}
\newcommand\xivec{{\boldsymbol \xi}}
\newcommand\etavec{{\boldsymbol \eta}}

\newcommand\Ubar{\overline{U}}
\newcommand\where{\hbox{\rm where\ }}
\newcommand\Ecalbar{\bar{\cal E}}
\newcommand\Tr{\hbox{\rm Tr}\,}
\newcommand{\rvec}{{\bf r}}
\newcommand{\svec}{{\bf s}}
\newcommand{\tvec}{{\bf t}}

\newcommand{\fvec}{{\bf f}}
\newcommand{\ehat}{{\bf \hat e}}
\newcommand{\Svec}{{\bf S}}
\newcommand{\Rr}{{\mathbb R}}
\newcommand{\Cc}{{\mathbb C}}
\newcommand{\Cbar}{{\overline C}}
\newcommand{\Ebarcal}{\overline{\Ecal}}
\newcommand\bra[1]{\left\langle #1 \right|}
\newcommand\ket[1]{{\left| #1 \right\rangle}}
\newcommand\sgn{\,{\hbox{\rm sgn}}}
\newsymbol \ltimes 226E
\newsymbol \rtimes 226F
\newcommand\braket[2]{\left\langle #1 \right.\left|#2\right\rangle}
\newcommand\rvecop{{\bf r}^{op}}
\newcommand\pvecop{{\bf p}^{op}}
\newcommand\svecop{{\bf s}^{op}}
\newcommand\rop{r^{op}}
\newcommand\pop{p^{op}}
\newcommand\Pop{P^{op}}
\newcommand\sop{s^{op}}
\newcommand\Hcal{{\cal H}}

\newcommand\jth{j^{th}}
\newcommand\rth{r^{th}}
\newcommand\Vcal{{\cal V}}
\newcommand\Ecal{{\cal E}}
\newcommand\Qsl{Q^{s\lambda}}
\newcommand\Vsl{\Vcal^{s\lambda}}
\newcommand\Esl{\Ecal^{s\lambda}}

\newcommand\Null{\hbox{\rm Nul}}
\newcommand\Spn{\hbox{\rm Spn}}
\newcommand\defn{\stackrel{\rm def}{\mathop=}}

\newcommand\Exp{\hbox{\rm Exp}\,}

\newcommand\ad{\hbox{\rm ad\,}}
\newcommand{\falg}{{\mathfrak f}}
\newcommand{\galg}{{\mathfrak g}}
\newcommand{\talg}{{\mathfrak t}}
\newcommand{\talgperp}{\talg^\perp}

\newcommand{\slm}{{{s\lambda}}}

\begin{document}

\thispagestyle{empty}

%\noindent ULM-TP/03-1\\ February 2003 \vspace*{1.5cm}

\begin{center}

{\LARGE\bf Quantum indistinguishability from\\
\vspace*{2mm}  
general representations of $SU(2n)$}\\ 
\vspace*{0.5cm}
{\large JM Harrison}%
\footnote{E-mail address: {\tt jon.harrison@physik.uni-ulm.de}}
$^{\dag,\ddag}$
{\large and JM Robbins}%
\footnote{E-mail address: {\tt j.robbins@bristol.ac.uk}}
$^\dag$

\vspace*{0.5cm}

$\dag$ School of Mathematics, University of Bristol, University Walk, 
Bristol BS8 1TW, UK\\
$\ddag$
Abteilung Theoretische Physik, 
Universit\"at Ulm, Albert-Einstein-Allee 11,\\ 
D-89069 Ulm, Germany 
\end{center}

\vfill

%\newpage
\begin{abstract}
A treatment of the spin-statistics relation in
  nonrelativistic quantum mechanics due to Berry and Robbins 
[{\it Proc. R. Soc. Lond.} A (1997) {\bf 453}, 1771-1790]
  is generalised within a group-theoretical framework.  The
  construction of Berry and Robbins is re-formulated in terms of
  certain locally flat vector bundles over $n$-particle configuration
  space.  It is shown how families of such bundles can be constructed
  from irreducible representations of the group $SU(2n)$.  The
  construction of Berry and Robbins, which leads to a definite
  connection between spin and statistics (the physically correct
  connection), is shown to correspond to the completely symmetric
  representations.  The spin-statistics connection is typically broken
  for general $SU(2n)$ representations, which may admit, for a given
  value of spin, both bose and fermi statistics, as well as
  parastatistics.  The determination of the allowed values of the spin
  and statistics reduces to the decomposition of certain zero-weight
  representations of a (generalised) Weyl group of $SU(2n)$.  A
  formula for this decomposition is obtained using the
  Littlewood-Richardson theorem for the decomposition of
  representations of $U(m + n)$ into representations of $U(m)\times
  U(n)$.
\end{abstract}

%\noindent PACS numbers: 03.65.-w, 03.65.Vf, 02.20.Qs

\newpage

\section{Introduction}\label{sec:intro}

In nonrelativistic quantum mechanics, the spin-statistics relation
specifies the behaviour of many-body wavefunctions for
indistinguishable particles under the exchange of a pair of particle
labels, and asserts that the wavefunctions either remain the same or
change sign according to whether the spin of the particles, $s$, is
integral or half-odd-integral.  Nonrelativistic quantum mechanics can
be formulated in a logically consistent way without the
spin-statistics relation, or else, with the wrong (ie, physically
incorrect) spin-statistics relation.  Therefore, if one is to derive
the spin-statistics relation from within a nonrelativistic theory, the
nonrelativistic theory must be reformulated, with postulates different
from the standard ones.  Whether such a reformulation serves to
explain the spin-statistics relation is, to some extent, a matter of
judgement, and depends on the naturalness and simplicity of the
assumptions introduced.

Such a reformulation was presented by Berry and Robbins 
%(1997)
\cite{paper:berryrobbins} (referred to in what follows as BR).  In BR,
the representation of spin was made to depend on position so that, in
contrast to the standard formulation, the $n$-particle wavefunction
was single-valued on configuration space.  The statistics of the
wavefunction was determined by a topological property of this
position-dependent spin representation.  A calculation showed that the
statistics were in accord with the physically correct spin-statistics
relation.  The construction was based on Schwinger's representation of
spin as number states of harmonic oscillators. Its implementation
assumed without proof the solution of a certain topological problem; a
solution was subsequently found by Atiyah 
%(2001) 
\cite{paper:atiyah1}.
The extension to relativistic wave equations was discussed by Anandan
\cite{paper:anandan}.

To be compelling, a derivation of the spin-statistics relation should
be based on general physical and mathematical principles, rather than
a particular construction.  In BR, it was suggested that certain
properties of the construction introduced therein might be sufficient
to ensure the correct spin-statistics relation.  Later, it was shown
that this is not the case 
%(Berry and Robbins 1999)
\cite{paper:berryrobbins2}), as alternative constructions exist which
possess these properties but yield the wrong statistics.  Thus, a 
nonrelativistic derivation of the spin-statistics relation
from general principles remains to be established along these lines.  For
a discussion of other nonstandard approaches to the spin-statistics
relation, see \cite{book:ducksudarshan, paper:ducksudarshan2}.
Our purpose here is to investigate a certain group-theoretical
generalisation of the construction in BR.  We begin in
Section~\ref{sec:bundle} by framing the underlying requirement, namely
that wavefunctions be single-valued, in a geometrical context.  The
setting for the quantum description of $n$ indistinguishable particles
are certain vector bundles over configuration space, which we call
$n$-spin bundles.  $n$-spin bundles carry a representation of the
spin-statistics group $\Sigma(n)$, which is (nearly) the group
generated by permutations and independent rotations of $n$ spinors
(the precise definition is given in Section~\ref{spins and perms}).
The particular representation of $\Sigma(n)$ characterises the spin
and statistics of the particles.  The statistics are then embodied in
a topological property of the $n$-spin bundle, namely the monodromy of
its flat connection.  This formulation is in the spirit of earlier
treatments by Leinaas \& Myrheim \cite{paper:leinaasmyrheim}
and Sorkin  \cite{paper:sorkin}.

In Section~\ref{sec:su2n} it is shown that $n$-spin bundles can be
constructed from irreducible representations $\Gamma^\fvec$ of the
group $SU(2n)$. The construction in BR is seen to be a particular
case, corresponding to the completely symmetric representations of
$SU(2n)$. For the completely symmetric representations, one obtains a
definite connection between spin and statistics, indeed the physically
correct connection. In contrast, an arbitrary representation of
$SU(2n)$ does not necessarily engender a definite relation between
spin and statistics; whether or not it does depends on the
decomposition of certain representations of the spin-statistics group
constructed from $\Gamma^\fvec$.

This decomposition is carried out in Section~\ref{sec:decomposition}.
The calculation involves the evaluation of integrals over characters
of the spin-statistics group, and makes use of the
Littlewood-Richardson formula for the decomposition of representations
of $U(k + l)$ into representations of $U(k)\ttimes U(l)$. It turns out
that for an arbitrary representation of $SU(2n)$ 
and a given value of spin, various choices of statistics may be
realised, including parastatistics (which correspond to
representations of the symmetric group of dimension greater than one).

Section~\ref{sec:discussion} contains a summary and discussion of the
results.  A connection to a more general problem in representation
theory is described in the Appendix.

Throughout this paper we will use the following notation: Given $n$
elements $a_1, \ldots, a_n$ of a set $\cal A$, we let $A$ denote the
ordered $n$-tuple $(a_1,\ldots, a_n$).  The action of a permutation
$\sigma \in S_n$ on $A$ is denoted by $\sigma\cdot A$, and defined by
\begin{equation}
  \label{eq:permutations of tuples}
  \sigma\cdot A = (a_{\sigma^{-1}(1)},\ldots,a_{\sigma^{-1}(n)}).
\end{equation}

Many of the results presented here are discussed in greater detail in
Harrison \cite{thesis:harrison}.

\section{Bundle description of $n$-particle quantum \\mechanics}
\label{sec:bundle}

The configuration space $C_n$ for $n$ particles in three-dimensional space
is the set of $n$-tuples $R = (\rvec_1,\ldots, \rvec_n)$.  We will
suppose the particles cannot coincide, so that $\rvec_j \ne \rvec_k$.
If the particles are indistinguishable, then permuted configurations
$R$ and $\sigma\cdot R$ are to be regarded as being the same.
  We describe here a framework for quantum mechanics in which
wavefunctions of identical particles are single-valued on
configuration space; that is, 
the wavefunction at
permuted configurations is the same. 

We first introduce in Section~\ref{spins and perms} the particular 
irreducible representations of the spin-statistics group $\Sigma(n)$,
denoted by $\Qsl$, which correspond to $n$ identical spins.
$n$-spin-$s$ bundles with statistics $\lambda$ are defined in
Section~\ref{subsec:bundles}.  These are flat, hermitian vector bundles
over configuration space whose fibres carry an irreducible
representation of the spin-statistics group equivalent to $\Qsl$.
This representation is required to be compatible with
indistinguishability and the flat connection. Wavefunctions are taken
to be sections of the bundle, and operators representing quantum
observables are defined on them. The relation to the standard formulation of
quantum mechanics, as well as that of BR, is discussed.

\subsection{Representations of the spin-statistics group for identical
  spinors}
\label{spins and perms} 

Let $S_n$ denote the symmetric group.  The irreducible
representations, $\Lambda^\lambda$, of $S_n$ are characterised by
Young tableaux, $\lambda$, of $n$ boxes (equivalently, partitions of
$n$).  Let $d_\lambda$ denote the dimension of the representation
$\Lambda^\lambda$.  Let $\ket{a}$, $a = 1,\ldots, d_\lambda$ denote an
orthonormal basis for $\Cc^{d_\lambda}$ (with respect to the standard
inner product).  For $\sigma \in S_n$, we write
\begin{equation}
  \label{eq:S_n irrep}
  \Lambda^\lambda(\sigma)\ket{a} = \Lambda^\lambda_{a',a}(\sigma)\ket{a'},
\end{equation}
where here and elsewhere a sum over repeated indices is implied.  We
may take $\Lambda^\lambda$ to be unitary, so that
$\Lambda^\lambda_{a',a}(\sigma)$ is a unitary matrix.

Let
\begin{equation}
  \label{eq:SU(2)^n}
  SU(2)^n = \underbrace{SU(2)\times\cdots
\times SU(2)}_{\hbox{$n$\ \rm times}}
\end{equation}
denote the direct product of $n$ copies of $SU(2)$. $SU(2)^n$
describes the independent rotations of $n$ spinors.  Denote elements
of $SU(2)^n$ by $U = (u_1,\ldots, u_n)$, with $u_j \in SU(2)$.  States
of $n$ spinors, all of spin $s$, are unchanged if pairs of spinors are
rotated through $2\pi$, regardless of whether $s$ is integral or
half-odd-integral.  Let $\Null(n) \subset SU(2)^n$ denote the subgroup
generated by pairs of $2\pi$-rotations.  It consists of
elements of the form
\begin{equation}
  \label{eq:null rotations}
U_0 = \left((-1)^{e_1} \UI_2, \ldots, (-1)^{e_n} \UI_2 \right), \quad \where
(-1)^{e_1}\cdots(-1)^{e_n} = 1
\end{equation}
($\UI_2$ is the $2\times 2$ identity matrix). The {\it $n$-spin group}, denoted by
$\Spn(n)$, is defined by
\begin{equation}
  \label{eq:n spin group}
  \Spn(n) = SU(2)^n/\Null(n),
\end{equation}
and represents in a one-to-one fashion the independent rotations of $n$ spinors of
the same spin.  Given $U \in SU(2)^n$, let
\begin{equation}
  \label{eq:[U]}
  \Ubar = U\, \Null(n)
\end{equation}
denote the corresponding element of $\Spn(n)$ (that is, $\Ubar$ is the
coset of $SU(2)^n$ containing elements which differ from $U$ by an
even number of $2\pi$ rotations).
%We remark that as far as our calculations are
%concerned, the distinction between $SU(2)^n$ and $\Spn(n)$ will not
%play a significant role.

The {\it spin-statistics group},
\begin{equation}
\label{eq:semidirect}
\Sigma(n) =  \Spn(n) \rtimes S_n,
\end{equation}
is the semidirect product of the $n$-spin group and the symmetric
group.  Elements are denoted by $(\Ubar,\sigma)$, where $U \in SU(2)^n$
and $\sigma \in S_n$, and multiplication is given by
\begin{equation}
  \label{eq:spin perm mult}
  (\Ubar,\sigma)(\overline{U'},\sigma') = (\overline{U (\sigma\cdot U'}),
 \sigma\sigma').
\end{equation}
(It is easy to check that the right-hand side of \eqref{eq:spin perm
mult} is unchanged if $U$ and $U'$ are multiplied by an even number of
$2\pi$ rotations.)  For brevity, when $\Spn(n)$ and $S_n$ are to be
regarded as subgroups of $\Sigma(n)$, we will denote their elements
simply by $\Ubar$ and $\sigma$ respectively, rather than by
$(\Ubar,\UI_{S_n})$ and $(\UI_{{\scriptstyle \rm Sp}(n)}, \sigma)$.

The complete set of irreducible representations of the spin-statistics
group can be obtained from the general representation theory of
semidirect products (see, eg, Mackey \cite{book:mackey}). 
Here we shall only be
interested in representations whose restriction to $\Spn(n)$ describes
$n$ spinors all of spin $s$, where $s$ is integral or
half-odd-integral.  It is easily established that each such
irreducible representation of $\Sigma(n)$ is characterised by $s$ and,
additionally, by an irreducible representation $\lambda$ of $S_n$.  We
denote this representation by $\Qsl$, and describe it in the
following.

$\Qsl$ acts on the  $(2s+1)^n d_\lambda$-dimensional vector
space $\Vsl$ given by
\begin{equation}
\label{eq:Vnslambda}
  \Vsl =
  \underbrace{\Cc^{2s+1}\otimes\cdots
\otimes\Cc^{2s+1}}_{\hbox{$n$\ \rm times}} \otimes \Cc^{d_\lambda}.
\end{equation}
Let
\begin{equation}
\label{eq:spinstat basis}
\ket{M, a} = \ket{m_1}\otimes\cdots\otimes\ket{m_n}\otimes \ket{a},
\end{equation}
where $M = (m_1,\ldots,m_n)$ and $m_j$ ranges between $-s$ and $s$ in
integer steps,
denote a
basis for $\Vsl$ orthonormal  with respect to
the standard inner products on $\Cc^{2s+1}$ and $\Cc^{d_\lambda}$.
For $\Ubar\in \Spn(n)$, $\Qsl(\Ubar)$ is given by
\begin{equation}
  \label{eq:Q(U)}
  \Qsl(\Ubar)\ket{M, a} = 
D^s_{m'_1,m_1}(u_1)\cdots D^s_{m'_n,m_n}
(u_n) \ket{M, a},
\end{equation}
where $D^s_{m,m'}(u)$ denotes the standard spin-$s$ representation of
$SU(2)$ on $\Cc^{2s+1}$, and $M' = (m'_1, \ldots, m'_n)$.  (It is
easy to check that the right-hand side of~\eqref{eq:Q(U)} is
unchanged if $U = (u_1,\ldots, u_n)$ 
is multiplied by an element of $\Null(n)$.)
For $\sigma \in S_n$, $\Qsl(\sigma)$ is given by
\begin{equation}
  \label{eq:Q(sigma)}
  \Qsl(\sigma)\ket{M, a}=
\Lambda^\lambda_{a',a}\ket{\sigma\cdot M, a'}.
\end{equation}
That is, the spin labels $M$ are permuted while $\ket{a}$
transforms according to the representation $\Lambda^\lambda$ of $S_n$.  For
a general element $(\Ubar,\sigma) \in \Sigma(n)$, the expression for
$\Qsl(\Ubar,\sigma)$ follows from \eqref{eq:Q(U)}
and \eqref{eq:Q(sigma)} and the multiplication law \eqref{eq:spin perm
  mult}.

\subsection{$n$-spin-$s$ bundles with statistics $\lambda$}\label{subsec:bundles}

For our purposes, a $k$-dimensional hermitian vector bundle, $\Ecal$,
over the configuration space $C_n$ will be regarded as a field of
$k$-dimensional subspaces, $\Ecal_R$, of a finite-dimensional Hilbert
space, $\Vcal$, depending smoothly on $R \in C_n$. $\Ecal_R$ is called the fibre
of $\Ecal$ at $R$.  A hermitian inner product on $\Ecal_R$ is induced by the
hermitian inner product on $\Vcal$.  A section of $\Ecal$ is a
function $\ket{\Psi(R)}$ on configuration space taking values in
$\Ecal_R$. The inner product of two sections is given by
\begin{equation}
  \label{eq:Hilbert space inner product}
  \int_{C_n} \braket{\Psi(R)}{\Phi(R)} \, dR.
\end{equation}
The space of square-integrable sections forms a Hilbert space.

To represent spin $s$ and statistics $\lambda$, each fibre $\Ecal_R$
must carry a representation of the spin-statistics group unitarily
equivalent to $\Qsl$.  Denote this representation by $L_R$.  We
require that $L_R$ depend smoothly on $R$.

Operators representing
spin, position and momentum may be defined on wavefunctions as
follows.  We consider the spin operators, 
denoted $\Svec^{op} = (\svecop_1,\ldots,\svecop_n)$, first.
Consider the
rotation of the $\rth$ spinor about an axis $\ehat_a$ by an angle $t$
holding the other spinors fixed. This is described by $U_{(r,a)}(t) =
(u_1(t),\ldots,u_n(t)) \in SU(2)^n$, where
\begin{equation}
  \label{eq:u_i(t)}
  u_j(t) = \begin{cases} \exp(-\ui t\sigma_a/2),& j = r\\
                         \UI_2,& {\rm otherwise},
           \end{cases}
\end{equation}
(here $\sigma_1$, $\sigma_2$, $\sigma_3$ are the Pauli matrices).
Then $\sop_{r,a}$, the $a$th component of $\svecop_r$, is given by
\begin{equation}
  \label{eq:spinop}
    \ket{\left(\sop_{r,a} \Psi\right)(R)} = \frac{1}{\ui}
\left.\frac{d}{d t}\right|_{t=0} 
L_R({\Ubar_{(r,a)}}(t))\, \ket{\Psi(R)}.
\end{equation}
As the representation $L_R$ is unitary, $\svecop_j$, as defined by
\eqref{eq:spinop}, is self-adjoint.  The representation property of
$L_R$ implies that the standard commutation relations for spin are
satisfied.
%, ie $[\sop_{j,a},\sop_{k,b}] = \ui \epsilon_{abc}
%\delta_{j,k} \sop_{j,c}.$
%\begin{equation}
%  \label{eq:spin commute}
%  [\sop_{j,a},\sop_{k,b}] = i
%  \epsilon_{abc} \delta_{j,k}
%  \sop_{j,c} 
%\end{equation}
%is satisfied. 

Position operators, $R^{op}=
(\rvecop_1,\ldots,\rvecop_n)$, are defined component-wise by
\begin{equation}
  \label{eq:rop}
  \ket{(\rop_{j,a} \Psi)(R)} = r_{j,a}
 \ket{\Psi(R)}.
\end{equation}
$\rvecop_j$ is hermitian with respect to the inner
product~\eqref{eq:Hilbert space inner product}, self-adjoint on a
suitable domain, and the position operators commute amongst each other
and with the spin operators.
%
%\begin{equation}
%  \label{eq:position commute}
%  [\rvecop_j,\rvecop_k] = 0, \quad [\rvecop_j, \svecop_k] = 0.
%\end{equation}

The definition of momentum operators requires a hermitian connection
on $\Ecal$.  A hermitian connection associates to piecewise smooth
paths $R(t) \in C_n$  a family of unitary maps between
the fibres $\Ecal_{R(t)}$.  These unitary maps describe the parallel
transport of spinors along $R(t)$.  Momentum operators may be defined
in terms of the covariant derivative with respect to this connection.
%, ie the rate of change of
%parallel-transported vectors.

A characteristic property of a connection is its curvature, which
describes parallel transport around infinitesimal closed paths.
Nonvanishing curvature corresponds physically to the presence of gauge
(eg, magnetic) fields.  In order that our theory be capable of
describing physics in the absence of fields, we shall require that
$\Ecal$ admit a flat connection.  This condition is not
automatically satisfied; the existence of a flat connection depends
on the topology of the bundle (just as the fact that a two-torus
admits a flat Riemannian metric, while a two-sphere does not, is a
consequence of their different Euler characteristics).

For a flat connection, parallel transport around a closed path is
trivial, provided the path is contractible.  In $C_n$, every closed
path is contractible ($C_n$ is simply connected).  Therefore,
parallel-transport with respect to a flat connection on $\Ecal$ is
path-independent, and depends only on the endpoints of the path.
Therefore, a flat hermitian connection on $\Ecal$ is characterised by
unitary maps $T_{R'\leftarrow R}: \Ecal_R \rightarrow \Ecal_{R'}$
describing parallel transport from $R$ to $R'$. Path
independence then implies that
\begin{equation}
  \label{eq:flat momentum}
  T_{R''\leftarrow R'} T_{R'\leftarrow R} = T_{R''\leftarrow R}.
\end{equation}

Momentum operators $\Pop = (\pvecop_1,\ldots,\pvecop_n)$ are defined
as follows. Let
\begin{equation}
  \label{eq:R_t}
  E_{(j,a)} = 
(0,\ldots,0,\ehat_a,0,\ldots,0) 
\end{equation}
denote the tangent vector in configuration space on which the $\jth$
particle moves with unit velocity in the direction $\ehat_a$ while the
other particles stay fixed. Then the $a$th component of $\pvecop_j$ is
given by
\begin{equation}
  \label{eq:momop}
  \ket{(\pop_{j,a} \Psi)(R)} = \left.\frac{d}{dt}\right|_0 
\left(
T_{R\leftarrow R+t E_{(j,a)}}  \ket{\Psi(R + t E_{(j,a)})} \right).
\end{equation} 
$\pop_{j,a}$ is hermitian with respect to the inner
product~\eqref{eq:Hilbert space inner product} and is self-adjoint on
a suitable domain. From~\eqref{eq:rop} it is easily verified that the
position and momentum operators satisfy the standard commutation
relations.
%, $[\rop_{j,a},\pop_{k,b}] = \ui \delta_{j,k} \delta_{a,b}$.
That the momentum operators commute amongst themselves follows from
the fact that
\begin{equation}
  \label{eq:momentum commute 2}
  T_{R\leftarrow R+tE} T_{R + tE \leftarrow R+tE
+ uF} = T_{R\leftarrow R + uF} T_{R + uF \leftarrow R + tE + uF},
\end{equation}
which in turn follows from the path independence~\eqref{eq:flat
  momentum} of the connection, provided the displacements $tE$ and
  $uF$ are small enough so as not to make the particles coincide.  

The requirement that spin and momentum commute is equivalent to
the requirement that parallel transport be compatible with the
representation $L_R$.  That is, we should have that
$L_{R'}(\Ubar,\sigma) T_{R'\leftarrow R} = T_{R'\leftarrow R}
L_{R}(\Ubar,\sigma)$.

%\begin{equation}
%    \label{eq:compatible}
%  L_{R'}(\Ubar,\sigma) T_{R'\leftarrow R} =  T_{R'\leftarrow R}
%  L_{R}(\Ubar,\sigma).
%\end{equation}
%If this is so, it is straightforward to show that
%$[\pvecop_j,\svecop_k] = 0$.

As a basis for subsequent discussion, let us formulate the standard
description of $n$-particle quantum mechanics within the framework
described above.  (In this case, the vector bundle description is
unnecessary, of course, and appears artificial.)
%In keeping with the
%notation in BR, we'll use the subscript $\std$ for the standard
%description (in BR, $F$ stood for ``fixed-basis wavefunctions'', in
%contrast to ``transported-basis wavefunctions'').
For this, take the fibres $\Ecal_{R
%,\std
}$ to be everywhere equal to the fixed vector space $\Vsl$.  Take
$L_{R
%,\std
}$, the representation of the
spin-statistics group, to be everywhere equal to the standard
representation $\Qsl$.  Parallel transport is everywhere taken to be
trivial; ie $T_{R'\leftarrow R
%,\std
}$ is just the identity map on
$\Vsl$.  Then $\Ecal$
%_\std$ 
is just the cartesian product $C_n \times
\Vsl$, and wavefunctions $\ket{\Psi
%_\std
(R)}$ are just $\Vsl$-valued
functions on $C_n$.  Wavefunctions may be expanded in the standard
basis $\ket{M,a}$ (cf~\eqref{eq:spinstat basis}),
\begin{equation}
  \label{eq:expand}
  \ket{\Psi(R)} = \sum_M \sum_{a = 1}^{d_\lambda} \psi_{M,a}(R) \ket{M,a},
\end{equation}
The definitions ~\eqref{eq:spinop},~\eqref{eq:rop}
and~\eqref{eq:momop} of the position, spin and momentum operators
yield the standard operations on the coefficients
$\psi_{M,a}(R)$,
\begin{eqnarray}
  \label{eq:position}
  \rvecop_j \psi^S_{M,a}(R) &=&  \rvec_j \psi^S_{M,a}(R),\\
\label{eq:momentum}
  \pvecop_j \psi^S_{M,a}(R) &=&  -\ui \nabla_{\rvec_j} \psi^S_{M,a}(R),
  \\
\label{eq:spin}
 \ue^{- \ui \theta \svecop_{j,a}}\psi^S_{M,a}(R) 
&=& 
\sum_{m' = -s}^s D^s_{m_j,m'}(\ue^{-\ui \theta\cdot \sigma_a})\psi^S_{M',a}(R),
\end{eqnarray}
where, in~\eqref{eq:spin}, $M'$ differs from $M$ only in the $j$th
component, in which $m_j$ is replaced by $m'$.  
%For another bundle,
%$\Ecal$, to yield a description physically equivalent to the standard
%one, there should be a one-to-one correspondence between $\Ecal$ and
%$\Ecal^\std$ preserving parallel-transport and the action of the
%spin-statistics group, and between the physically allowed sections of
%$\Ecal$ and $\Ecal^\std$.

We now introduce the requirement, basic to the formulation in BR, that
for indistinguishable particles, the values of the wavefunction at
permuted configurations should be the same.  That is, we require that
%$\ket{\Psi(\sigma\cdot R} = \ket{\Psi(R)}$.
\begin{equation}
  \label{eq:singlevalued 1}
  \ket{\Psi(\sigma\cdot R)} = \ket{\Psi(R)}, \quad \sigma \in S_n.
\end{equation}
(Note that for this condition to be sensible, the fibres at $R$ and
$\sigma\cdot R$ must be the same.)  In this case, the wavefunction is
single-valued as a function of configurations in which the particles
are no longer labeled.  
Wavefunctions in the standard description are
not single-valued in this sense.  Indeed, in the standard description,
the coefficients of the wavefunction at permuted configurations are
related by
\begin{equation}
  \label{eq:standard case: permutation}
  \psi_{M,a}(\sigma\cdot R)= \Lambda^{\lambda}_{a,a'} 
\psi_{\sigma^{-1}\cdot M, a'}(R),
\end{equation}
so that the wavefunctions themselves satisfy
\begin{equation}
\label{eq:??}
 \ket{\Psi(\sigma\cdot R)} = L_{\sigma\cdot R}(\sigma)
\ket{\Psi(R)}.
\end{equation}

Descriptions based on single-valued wavefunctions, but physically
equivalent to the standard description, are obtained by re-writing
\eqref{eq:??} as
\begin{equation}
  \label{eq:all cases: permutation}
    \ket{\Psi(\sigma\cdot R)} = L_{\sigma\cdot R}(\sigma)
  T_{\sigma\cdot R\leftarrow R}
\ket{\Psi(R)}.
\end{equation}
In the standard description, \eqref{eq:all cases: permutation} is the
same as \eqref{eq:??}, since $T_{\sigma\cdot R\leftarrow R}$ is just
the identity in this case.  In contrast, For single-valued wavefunctions,
\eqref{eq:all cases: permutation} becomes
\begin{equation}
  \label{eq:parallel transport + single valued}
  L^{-1}_{\sigma\cdot R} (\sigma) = T_{\sigma\cdot R\leftarrow R}.
\end{equation}
Thus, for a description in terms of single-valued wavefunctions to be
equivalent to the standard one, parallel transport is necessarily a
nontrivial operation; between permuted configurations, parallel
transport induces the corresponding permutation of spins.  
%The
%formalism of vector bundles, while contrived and artificial in the
%standard description, is the appropriate language for describing this
%phenomenon, as already discussed lucidly by Leinaas \& Myrheim.

Let us now formalise the preceding considerations.  An $n$-spin-$s$
bundle with statistics $\lambda$,  denoted by $\Ecal^\slm$, is defined to be
a $(2s+1)d_\lambda$-dimensional hermitian vector bundle over the
configuration space $C_n$ endowed with the following properties:
\renewcommand{\labelenumi}{\Alph{enumi})}
\begin{enumerate}
\parindent = -12 pt
\item There exists a smooth family, $L_R$, of unitary irreducible
  representations of the spin-statistics group $\Sigma(n)$ acting on
  the fibres $\Ecal_R$, unitarily equivalent to $\Qsl$.
  \label{it:repns}
\item The fibres at permuted configurations are the same, ie
\begin{equation}
  \label{eq:fibres same}
  \Ecal_{\sigma\cdot R} = \Ecal_R.
\end{equation}\label{it:same fibres}

\item There exists a flat hermitian connection on $\Ecal^\slm$,
  characterised by unitary maps $T_{R'\leftarrow R}$ describing
  parallel transport from $R$ to $R'$, satisfying the composition rule
\begin{equation}
  \label{eq:flat momentum 2}
  T_{R''\leftarrow R'} T_{R'\leftarrow R} = T_{R''\leftarrow R}.
\end{equation}
Parallel transport is compatible with the representation $L_R$ in the
sense that
\begin{equation}
    \label{eq:compatible 2}
  L_{R'}(\Ubar,\sigma) T_{R'\leftarrow R} =  T_{R'\leftarrow R}
  L_{R}(\Ubar,\sigma).
\end{equation}

\item Parallel transport between permuted fibres induces permutations, ie
\begin{equation}
  \label{eq:monodromy}
  T_{\sigma\cdot R\leftarrow R} = L_{\sigma\cdot R}(\sigma^{-1}).
\end{equation}
\end{enumerate}
%\parindent = \saveparindent

The Hilbert space $\Hcal$ of wavefunctions describing $n$ indistinguishable
particles of spin $s$ and statistics $\lambda$ is the space of sections
of $\Esl$ with inner product~\eqref{eq:Hilbert space inner product}
satisfying the single-valuedness condition
\begin{equation}
 \label{eq:singlevalued}
 \ket{\Psi(\sigma\cdot R)} = \ket{\Psi(R)}.
\end{equation}
Observables are generated by combinations of
the position, momentum and spin operators, $\rvecop_j$, $\pvecop_j$
and $\svecop_j$, given by~\eqref{eq:rop},~\eqref{eq:momentum}
and~\eqref{eq:spinop} respectively, which are invariant under
permutations. These permutation-invariant operators preserve the
single-valuedness condition \eqref{eq:singlevalued}.

To establish explicitly the equivalence between this formulation and
the standard one, as well as the treatment in BR, it is useful to
introduce a parallel-transported basis for the fibres $\Ecal_R$.  To
this end, we fix a reference configuration $R_0 \in C_n$. Since
$L_{R_0}$ is unitarily equivalent to $\Qsl$, there exists an
orthonormal basis $\ket{M,a(R_0)}$ of $\Ecal_{R_0}$ for which
\begin{equation}
  \label{eq:reference basis}
  L_{R_0}(\Ubar,\sigma) \ket{M,a(R_0)} = \Qsl_{M'a',Ma}(\Ubar,\sigma)
 \ket{M',a'(R_0)}.
\end{equation}
A basis for $\Ecal_R$ is defined via parallel transport as follows:
\begin{equation}
  \label{eq:transported basis}
  \ket{M,a(R)} = T_{R\leftarrow R_0}  \ket{M,a(R_0)}.
\end{equation}
Because the representation $L_R$ is compatible with the flat
connection, it follows that
\eqref{eq:reference basis} holds for all $R$.
%, namely
%\begin{equation}
%  \label{eq: basis elsewhere}
%  L_{R}(U,\sigma) \ket{M,a}(R) = Q^{ns\lambda}_{M'a',Ma}
% \ket{M',a'}(R).
%\end{equation}

Wavefunctions $\ket{\Psi(R)}$ may be expanded in terms of this basis
as
\begin{equation}
  \label{eq:expand 2}
  \ket{\Psi (R)} = \psi_{M,a}(R) \ket{ M , a (R)}.
\end{equation}
From the definitions \eqref{eq:rop}, \eqref{eq:momop}  and
\eqref{eq:spinop}, it is readily verified that the position, momentum and
spins operators act on the components $\psi_{M,a}(R)$ as the
standard operators~\eqref{eq:position} -- \eqref{eq:spin}.  The
condition \eqref{eq:monodromy} implies that the components at permuted
configurations are related as in~\eqref{eq:standard case:
  permutation}, in accord with the standard formulation.

Apart from allowing parastatistics, the framework described here is
equivalent to the one given in Section~2 of BR. There are, however,
some differences in the formulation.  In BR, properties A) -- D) are
expressed directly in terms of the parallel-transported basis.  For
example, instead of property C), BR require that the
parallel-transported basis satisfy
\begin{equation}
  \label{eq:BR parallel}
  \braket{M'(R)}{\nabla_{\rvec_j}M(R)} = 0.
\end{equation}
In this way, the formalism and
terminology of vector bundles is avoided.

An advantage of the present formulation is that properties required by
physical considerations are distinguished from those which depend on
convention.  For example, \eqref{eq:BR parallel} implies the existence
of a flat connection, but it implies, additionally, that it is a
particular connection which is flat- namely, the connection induced by
the inner product on $\Vcal$, according to which vectors are
parallel-transported by translating them to an infinitesimally
displaced fibre and there projecting them perpendicularly. This
choice of connection, while convenient, is nevertheless a matter of
convention, and is not required by physical considerations.

Finally, we note that we could, if we wished, impose the
single-valuedness condition more directly by taking $n$-particle
configuration space to be the identified configuration space $\Cbar_n
= C_n/S_n$ consisting of (unordered) sets $X = \{\rvec, \svec,
\ldots, \tvec\}$ of $n$ distinct points in $\Rr^3$.  (This is the
point of view taken by Leinaas and Myrheim (1977).) 
%The preceding
%formulation can be recast in this way.  
Then wavefunctions would become functions
of $X$, or, more precisely, sections of an $n$-spin bundle over the
identified configuration space $\Cbar_n$.  However, such a
reformulation involves some additional mathematical complication, and,
for this reason, we will confine our consideration of it to the
following informal remarks.

The complication is due to the fact that there are no global Euclidean
coordinates on $\Cbar_n$; it is no longer sensible to refer to
position, spin and momentum operators for a particular particle.  
In place of individual momenta, for example, one must introduce
generalised momentum operators, which are related to covariant
derivatives along smooth vector fields on $\Cbar_n$. 
A formulation in terms of $\Cbar_n$ does have some attractive aspects,
though.  Parallel transport between permuted fibres in $C_n$ becomes
transport from a single fibre to itself around a non-contractible
closed path in $\Cbar_n$.  In this way, the statistics of the
particles is reflected in the monodromy of the flat connection, a
topological property of the bundle.

\section{Spin bundles $SU(2n)$ representations }\label{sec:su2n}

In this section we describe the construction of $n$-spin bundles from
representations of $SU(2n)$.
The construction is based on a connection
between $SU(2n)$ and the spin-statistics group $\Sigma(n)$
(Section~\ref{sec:su2npart1}), which associates 
a representation $\Delta^\fvec$ of
$\Sigma(n)$ to
an irreducible representation
$\Gamma^\fvec$ of $SU(2n)$
(Section \ref{sec:representations of spinstatistics,
su2n}).  In general, the representation $\Delta^\fvec$ is reducible.
$n$-spin bundles are constructed from the representations
$\Gamma^\fvec$ and $\Delta^\fvec$ and an $S_n$-equivariant map
from $C_n$ to $SU(n)/T(n)$ (Section~\ref{sec:construct spin bundles}).

Whether $\Gamma^\fvec$ determines a spin-statistics relation is
discussed in Section~\ref{sec:spin-stat relations}.  The question is
related to the decomposition of $\Delta^\fvec$ into its irreducible
components.  A definite statistics for a given value of spin
requires that $\Delta^\fvec$ should contain only one
irreducible representation of $\Sigma(n)$ with that spin. This is the
case for the completely symmetric representations, which correspond to
the construction in BR.  The general case is discussed in
Section~\ref{sec:decomposition}.

\subsection{The spin-statistics group and $SU(2n)$}\label{sec:su2npart1}

Consider $SU(2n)$, the group of $2n$-dimensional unitary matrices of
unit determinant.  
%When using indices, we will denote the components of
%$f\in SU(2n)$ by $f_{\ar,\bt}$, where $1 \le \alpha,\beta \le 2$ and
%$1 \le r,t \le n$.  This choice of indices reflects the factorisation
%$\Cc^{2n} = \Cc^2 \otimes \Cc^n$.  $\alpha$ and $\beta$ correspond to
%spin, while $r$ and $t$ correspond to particle labels.
%
$SU(2)^n$ may be identified as a subgroup of $SU(2n)$, with 
$U = (u_1,\cdots,u_n)\in SU(2)^n$ identified with the matrix
\begin{equation}\label{eq: U in SU(2)^n}
U = 
\left(
\begin{array}{cccc}
u_1 &     &        &     \\
    & u_2 &        & 0   \\
    &     & \ddots &     \\
    & 0   &        & u_n \\
\end{array}
\right).
\end{equation}
(For simplicity, we'll use the same symbol, in this instance $U$, for
both an element of $SU(2)^n$ and for the corresponding matrix in
$SU(2n)$, and will do the same for some other subgroups of
$SU(2n)$ to be introduced below.  Taken in context 
this usage should not introduce
any ambiguity).  Similarly, $SU(n)$, the group of $n$-dimensional
unitary matrices with unit determinant, may be identified with a
subgroup of $SU(2n)$, with $g \in SU(n)$, with components
denoted by $g_{rt}$, $1 \le r,t \le t$,
identified with the $SU(2n)$-matrix
\begin{equation}\label{eq: g in SU(n)}
g = 
\left(
\begin{array}{ccc}
g_{11}\UI_2   &\ldots   &g_{1n} \UI_2 \\
\vdots         &\ddots   &   \vdots   \\
g_{n1}\UI_2   &\ldots   &g_{nn} \UI_2 \\
\end{array}
\right).
\end{equation}
Finally, we let $T(n)$ denote the subgroup of diagonal matrices in
$SU(n)$.  From \eqref{eq: g in SU(n)},
$T(n)$ may be identified as the subgroup of $SU(2n)$
consisting of diagonal matrices of the form
\begin{equation}\label{eq: t in T(n)}
t(\Theta) = 
\left(
\begin{array}{cccc}
\ue^{\ui\theta_1}\UI_2 & & & \\
 & \ue^{\ui\theta_2}\UI_2 & & 0 \\
 & & \ddots & \\
 & 0 & & \ue^{\ui\theta_n}\UI_2 \\
\end{array}
\right),
\end{equation}
where $\Theta = (\theta_1,\ldots,\theta_n)$ is an $n$-tuple of phases
satisfying
\begin{equation}
  \label{eq:theta condition}
  \ue^{\ui \theta_1}\cdots \ue^{\ui \theta_n} = 1.
\end{equation}
Note that $SU(2)^n\cap T(n)$ is just the subgroup $\Null(n)$ of null
rotations, which consists of $SU(2n)$-matrices of the form
\begin{equation}\label{eq: U_0 in Nul(n)}
U_0 = 
\left(
\begin{array}{cccc}
(-1)^{e_1}\UI_2 & & & \\ 
 & (-1)^{e_2}\UI_2 & & \\ 
 & & \ddots & \\ 
 & & & (-1)^{e_n}\UI_2 \\
\end{array}
\right),
\end{equation}
where $(-1)^{e_1}\cdots(-1)^{e_n} = 1$.

Let $N(n) \subset SU(n)$ denote the normaliser of $T(n)$ in $SU(n)$,
ie the subgroup of $ SU(n)$ which leaves $T(n)$ invariant
under conjugation.  It is straightforward to show that elements of
$N(n)$ may be parameterised by a permutation $\sigma \in S_n$ and an
$n$-tuple of phases $\Phi = (\phi_1,\ldots,\phi_n)$ satisfying
\begin{equation}
  \label{eq:Phi condition} \ue^{\ui \phi_1}\cdots \ue^{\ui \phi_n} =
  \sgn(\sigma),
\end{equation}
(here $\sgn(\sigma)$ denotes the parity of $\sigma$), and are of the form
\begin{equation}
  \label{eq:y(sigma,phi)}
  y_{rt}(\sigma,\Phi) = \delta_{r,\sigma(t)} \ue^{\ui \phi_t}.
\end{equation}
Multiplication in $N(n)$ is given by $y(\sigma,\Phi) y(\sigma',\Phi')
= y(\sigma\sigma', \sigma'^{-1}\cdot \Phi + \Phi')$, so that, formally,
$N(n)$ may be regarded as the semidirect product, $S_n\rtimes T(n)$.
The quotient $N(n)/T(n)$, the Weyl group of $SU(n)$, is
isomorphic to $S_n$.

Let $M(n)$ denote the normaliser of $T(n)$ in $SU(2n)$, ie the
subgroup of $SU(2n)$ which leaves $T(n)$ invariant under conjugation.
Clearly $M(n)$ contains $N(n)$ as a subgroup.  $M(n)$ also contains 
the $SU(2n)$-centraliser of $T(n)$, denoted by $Z(n)$, ie the subgroup of
$SU(2n)$ whose elements commute with all elements of $T(n)$.  
It is straightforward to show that elements of $Z(n)$ are
of the form $U\, t(\Theta)$, where $U \in SU(2)^n$.
%is given by \eqref{eq: U
%in SU(2)^n}, 
and $t(\Theta)\in T(n)$.
%is given by \eqref{eq: t in
%T(n)}.  
It is then straightforward to show that elements of $M(n)$ can be
expressed as products of elements of $Z(n)$ and $N(n)$, and thus are
of the form
\begin{equation}
  \label{eq:param x}
  x(U,\sigma,\Phi) = U y(\sigma,\Phi),
\end{equation}
where the phases $\Phi$ satisfy \eqref{eq:Phi condition}. Multiplication 
in $M(n)$ is
given by 
\begin{equation}\label{eq:mult in M(n)}
x(U,\sigma,\Phi)\, x(U',\sigma',\Phi') = x(U\sigma\cdot U',
\sigma\sigma', \sigma'^{-1} \cdot \Phi + \Phi').
\end{equation}

The
parameterisation $x(U,\sigma,\Phi)$ of \eqref{eq:param x} is not unique. 
If $U$ is replaced
by $U U_0$, with $U_0\in \Null(n)$ given by \eqref{eq: U_0 in Nul(n)},
and $\Phi$ is replaced by $\Phi'$, where $\phi'_j = \phi_j + e_j \pi$,
then $x(U,\sigma,\Phi)$ is unchanged.  In this way, we
see that, formally, $M(n)$ is isomorphic to 
$SU(2)^n\ltimes N(n)/\Null(n)$.

From these considerations, it follows that the quotient
$M(n)/T(n)$
is
isomorphic to the spin-statistics group, ie
\begin{equation}
  \label{eq:M(n)/T(n) iso}
\Sigma(n) = \Spn(n) \rtimes S_n \cong M(n)/T(n).
\end{equation}
The isomorphism is given explicitly by 
\begin{equation}
\label{eq: iso explicit}
(\Ubar,\sigma)\mapsto
x(U,\sigma,\Phi) T(n),
\end{equation}
where $x(U,\sigma,\Phi) T(n)$ denotes a coset
in $M(n)/T(n)$.  This association between $SU(2n)$ and the
spin-statistics group is the basis of the constructions to follow.

\subsection{Representations of the spin-statistics group from 
representations of $SU(2n)$}
\label{sec:representations of spinstatistics,  su2n}

Let $\Gamma^\fvec$ denote a unitary irreducible representation of
$SU(2n)$, labeled by a Young tableau $\fvec = (f_1,\ldots,f_{2n})$ of
up to $2n$ rows (in fact, the last row of $\fvec$ may be taken to be
empty).  Let $\Vcal$ denote the hermitian inner product space on which
$\Gamma^\fvec$ acts.  (Of course, $\Vcal$ depends on the choice of
representation, but to simplify the notation we will not indicate this
explicitly.)

Under the restriction of $\Gamma^\fvec$ to $T(n)$, $\Vcal$ may be
decomposed into a direct sum of orthogonal subspaces, $\Vcal^{K}$, on
which $\Gamma^\fvec(t(\Theta))$ is represented 
by the phase factor $\exp \ui (K\cdot \Theta)$.  Here $K =
(k_1,\ldots, k_n)$ is an $n$-tuple of integers, and
$K\cdot\Theta = \sum_j k_j \theta_j$.  The subspace $\Vcal^{0}$,
corresponding to $K = (0,\ldots,0)$, consists of vectors which are
invariant under $\Gamma^\fvec(T(n))$.

Let us determine the action of $M(n)$ on the subspaces $\Vcal^K$.
Given $x(U,\Phi,\sigma) \in M(n)$, it follows from \eqref{eq:mult in
M(n)} that
\begin{eqnarray} 
  \label{eq:weight space 1}
 \Gamma^\fvec(t(\Theta)) \Gamma^\fvec(x(U,\sigma,\Phi))
   \cdot 
\Vcal^K
&=& \Gamma^\fvec(x(U,\sigma,\Phi))
 \Gamma^\fvec(t(\sigma^{-1}\cdot \Theta)) \cdot\Vcal^K\nonumber\\
&=& \ue^{\ui (K\cdot (\sigma^{-1} \cdot
   \Theta))} \Gamma^\fvec(x(U,\sigma,\Phi))
\cdot \Vcal^K\nonumber\\
&=& \ue^{\ui (\sigma\cdot K)\cdot
   \Theta}\Gamma^\fvec(x(U,\sigma,\Phi)) \cdot \Vcal^K.
\end{eqnarray}
Thus, under the action  of $M(n)$, the subspaces
$\Vcal^K$ are mapped into one another according to
\begin{equation}
  \label{eq:normaliser on weights}
\Gamma^\fvec(x(U,\Phi,\sigma)) \cdot \Vcal^K = V^{\sigma\cdot K}.
\end{equation}

It follows from \eqref{eq:normaliser on weights} that $\Vcal^0$ is
invariant under $M(n)$.  Therefore $\Gamma^\fvec$ restricts to a
representation of $M(n)$ on $\Vcal^0$. Since $T(n) \subset M(n)$
belongs to the kernel of this representation (as $T(n)$ leaves vectors
in $\Vcal^0$ invariant), $\Gamma^\fvec(M(n))$ reduces to a
representation of the quotient $M(n)/T(n)$, which we denote by
$\Delta^\fvec$.  Since $M(n)/T(n) \cong \Sigma(n)$ (cf
\eqref{eq:M(n)/T(n) iso}), $\Delta^\fvec$ is in fact a representation
of the spin-statistics group.  From \eqref{eq: iso explicit},
$\Delta^\fvec$ is given by
\begin{equation}\label{eq:Delta}
  \Delta^\fvec(\Ubar,\sigma) = \Gamma^\fvec(x(U,\Phi,\sigma)).
\end{equation}

In general, the representation $\Delta^\fvec$ of $\Sigma(n)$ is
reducible.  Let $\nu(\fvec,s\lambda)$ denote the multiplicity with
which the irreducible representation $\Qsl$, given by {eq:Q(U)},
appears in the decomposition of $\Delta^\fvec$.  This multiplicity
will play a central role in what follows.

\subsection{Construction of $n$-spin bundles}\label{sec:construct spin bundles}

Let $\Xi: C_n \rightarrow SU(n)/T(n)$ denote a smooth map from
$n$-particle configuration space $C_n$ to the coset space
$SU(n)/T(n)$. Such a map may be represented by $g(R)$, an
$SU(n)$-valued function on $C_n$ which is smooth up to right
multiplication by an element of $T(n)$.  (That is, discontinuities in
$g(R)$ can be removed locally by multiplying on the right by a
discontinuous $T(n)$-valued function.)  The symmetric group $S_n$ acts
on $C_n$ as permutations (ie, $R \mapsto \sigma\cdot R$) and on
$SU(n)/T(n)$ as the Weyl group (ie, for $y(\sigma,\Phi)\in N(n)$, $g\,
T(n) \mapsto g y^{-1}(\sigma,\Phi) \, T(n)$).  $\Xi$ is said to be
equivariant with respect to $S_n$ if, for all $\sigma \in S_n$, $\Xi
\circ \sigma = \sigma \circ \Xi$.  In terms of $g(R)$,
$S_n$-equivariance is equivalent to
\begin{equation}
  \label{eq:equivariance}
  g(\sigma \cdot R) \, T(n)  = g(R) y^{-1}(\sigma,\Phi) \, T(n).
\end{equation}
Atiyah (2000) has shown that there exist continuous (and therefore
smooth) $S_n$-equivariant maps from $C_n$ to $SU(n)/T(n)$.  
Let $g(R)$ represent any such equivariant map (the results
which follow do not depend on the particular choice of $g(R)$).

As in
Section~\ref{sec:representations of
spinstatistics, su2n}, let $\Gamma^\fvec$ be an irreducible representation
of $SU(2n)$, and $\Delta^\fvec$ the associated representation of 
the spin-statistics group $\Sigma(n)$. 
Suppose $\nu(\fvec,s\lambda) > 0$, ie
$\Vcal^0$ contains a subspace, which we denote by $\Vsl$,
which transforms under $\Delta^\fvec$ 
according to the irreducible representation $\Qsl$ of $\Sigma(n)$.
We may then
construct an $n$-spin-$s$ bundle $\Esl$ as follows.  
The fibres $\Esl_R \subset \Vcal$ 
are given by
\begin{equation}
  \label{eq:weight fibres}
  \Esl_R = \Gamma^\fvec(g(R)) \cdot \Vsl.
\end{equation}
Since $g(R)$ is smooth up to right multiplication by a $T(n)$-valued
function and $\Vsl$ is invariant under $\Gamma^\fvec(T(n))$, it
follows that $\Esl_R$ depends smoothly on $R$.

Let us verify that $\Esl$ has the properties A) -- D) listed in
Section~\ref{subsec:bundles}.  For A), we define the representation
$L_R$ on $\Esl_R$ by \begin{equation} \label{eq:defn of L_R}
L_R(\Ubar,\sigma) = \Gamma^\fvec(g(R)) \Delta^\fvec(\Ubar,\sigma)
{\Gamma^\fvec}^\dag(g(R)), \end{equation} where
$\Delta^\fvec(\Ubar,\sigma)$ is the representation of $\Sigma(n)$
given by \eqref{eq:Delta}.  By assumption, $\Delta^\fvec$ is unitarily
equivalent to $\Qsl$ on $\Vsl$, so it is evident from \eqref{eq:defn
of L_R} that $L_R$ is unitarily equivalent to $\Qsl$ for all $R$.
Since the right-hand side of \eqref{eq:defn of L_R} is unchanged if
$g(R)$ is multiplied on the right by a (possibly discontinuous)
$T(n)$-valued function, it is clear that $L_R$ depends smoothly on
$R$.

For B), from the definition \eqref{eq:weight fibres}
and the equivariance property \eqref{eq:equivariance}, we have that
\begin{eqnarray}
  \label{eq:fibres are the same}
  \Esl_{\sigma\cdot R} &=& 
 \Gamma^\fvec(g(\sigma \cdot R)) \cdot \Vsl\nonumber\\
&=& \Gamma^\fvec(g(R))\Gamma^\fvec(y^{-1}(\sigma,\Phi))\cdot \Vsl =
 \Gamma^\fvec(g(R))\cdot \Vsl \nonumber\\
&=& \Esl_R.
\end{eqnarray}
Thus the fibres at permuted configurations are the same.

For C), we define the unitary maps $T_{R'\leftarrow R}$ describing
flat parallel transport between the fibres at $R$ and $R'$ by
\begin{equation} \label{eq:parallel - construction} T_{R' \leftarrow
R} = \Gamma^\fvec(g(R')) {\Gamma^\fvec}^\dag(g(R)) \end{equation} The
right-hand side is unchanged if $g(R)$ is multiplied on the right by a
$T(n)$-valued function, so $T_{R' \leftarrow R}$ is well defined and
depends smoothly on $R$ and $R'$. The composition law \eqref{eq:flat
momentum 2} is easily verified.  Compatibility with the
representations $L_R$ (cf \eqref{eq:compatible 2}) follows from the
definition of $L_R$ in \eqref{eq:defn of L_R} and the representation
property of $\Gamma^\fvec$.
  
For D), from \eqref{eq:parallel - construction}, parallel transport
  $T_{\sigma\cdot R\leftarrow R}$ between permuted fibres $R$ and
  $\sigma\cdot R$ is given by $\Gamma^\fvec(g(\sigma \cdot R))
  {\Gamma^\fvec}^\dag(g(R))$.  The equivariance condition
  \eqref{eq:equivariance} implies this is equal to
  $\Gamma^\fvec(g(R))\Gamma^\fvec(y^{-1}(\sigma,\Phi))
  {\Gamma^\fvec}^\dag(g(R))$.  From \eqref{eq:defn of L_R}, the
  condition \eqref{eq:monodromy} follows.

\subsection{Spin-statistics relations from $SU(2n)$ representations?}\label{sec:spin-stat relations}

%what if f is completely symmetric representation.  calculation that
%gives m = ...  polynomials of degree y_1{s_1 + m_1} z_1(s_1 - m_1)
%all %s's the same gives ...  calculate sign again.

Given an irreducible representation $\Gamma^\fvec$ of $SU(2n)$, the
preceding construction determines the statistics for spin $s$
unambiguously, provided that there is just one representation $\Qsl$
with spin $s$ in the decomposition of $\Delta^\fvec$; equivalently,
given $\fvec$ and $s$, the multiplicity $\nu(\fvec,s,\lambda)$ should
vanish for all but one $\lambda$.

This is the case for the completely symmetric representations of
$SU(2n)$.  The completely symmetric representations correspond to
Young tableaux with a single row.  Let $\Gamma^d$ denote the 
representation for a single row of $d$ boxes.
$\Gamma^d$ may be
realised on the space $\Vcal$ of homogeneous polynomials of degree $d$
in $2n$ variables, $z = (z_1,\ldots,z_{2n}) \in \Cc^{2n}$, and is 
given by $\Gamma^d(f)\cdot P(z) = P(f^{-1}\cdot z)$ for $f\in
SU(2n)$.  $\Gamma^d$ is unitary with respect to the inner product
\begin{equation}
  \label{eq:inner product on P(z)}
  \langle P, Q \rangle = \int_{\Cc^{2n}} \ue^ {-{z^*\cdot z}/2} P^*(z)
  Q(z)\, d^{4n} z
\end{equation}
on $\Vcal$.

An orthogonal basis for $\Vcal$ is given by the monomials
\begin{equation}
  \label{eq:Pab}
  \prod_{r = 1}^n z_{1,r}^{a_r} z_{2,r}^{b_r},
\end{equation}
where the sum of the exponents $a_j$ and $b_j$ is given by $d$.  The
subspace $\Vcal^0$, whose vectors are invariant under $T(n)$, consists
of polynomials which are invariant under $z \mapsto
t^{-1}(\Theta)\cdot z$, where
\begin{equation}
 t^{-1}(\Theta)\cdot z = \left(\ue^{-\ui \theta_1} z_1, \ue^{-\ui \theta_1} z_2,
\ue^{-\ui \theta_n} z_{2n-1}, \ue^{-\ui \theta_n} z_{2n}\right)
\end{equation}
Such polynomials are linear combinations of the
monomials \eqref{eq:Pab} for which $a_r + b_r$ is independent of $r$,
so that $a_r + b_r = d/n$.  Thus, for $\Vcal^0$ to be nontrivial, $d$ must
be divisible by $n$.  

Let us assume this is the case, so that $d/n$ is integral.
Then $s = d/2n$ is either
integral or half-odd-integral.  Let $m_r = a_r - s$.  Then $a_r = s +
m_r$ and $b_r = s - m_r$.  It follows that $\Vcal^0$ is spanned by the
$(2s+1)^n$ monomials
\begin{equation}
  \label{eq:monomials}
   \prod_{r = 1}^n z_{1,r}^{s+m_r} z_{2,r}^{s-m_r},
\end{equation}
where $-s \le m_r \le s$ and $s \pm m_r$ is integral.  Under $\Spn(n)$,
$\Vcal^0$ transforms as $n$ spin-$s$ spinors.  As the dimension of
$\Vcal^0$ is $(2s+1)^n$, it follows that there is a single irreducible
representation $\Qsl$ with multiplicity one in the decomposition of
$\Delta^\fvec$, and that $d_\lambda = 1$, ie $\lambda$ is either the
completely symmetric or the completely antisymmetric representation of
$S_n$.

$\lambda$ may be determined by considering the action of permutations
on an element of $\Vcal^0$, for example 
\begin{equation}
P_s(z) \defn z_{2}^{2s}
z_{4}^{2s}\ldots z_{2n}^{2s}.
\end{equation}
From \eqref{eq:y(sigma,phi)}, under
$y^{-1}(\sigma,\Phi)$, the even components transform as  $z_{2j}
\mapsto \ue^{-\ui \phi_j} z_{2\sigma(j)}$.  Thus, under
$y(\sigma,\Phi)$, $P_s(z)$ is multiplied by the phase factor
$\ue^{2s\ui \phi_1}\cdots \ue^{2s\ui \phi_n}$.  From \eqref{eq:Phi
condition}, this phase factor is just $\sgn^{2s}(\sigma)$.  Thus, the
completely symmetric representations of $SU(2n)$ of dimension $d =
2ns$ lead to $n$-spin-bundles with spin $s$ and bose or fermi
statistics according to the parity of $2s$, in accord with the
physically correct spin-statistics relation. This is
precisely the result obtained in BR.

\section{Calculation of the multiplicities}\label{sec:decomposition}

Given an arbitrary representation $\Gamma^\fvec$ of $SU(2n)$, we wish to determine
for which spins an $n$-spin-$s$ bundle can be constructed, and, for
those spins, whether the constructions have a definite type of
statistics.  To this end, we calculate the multiplicities
$\nu(\fvec,\slm)$ with which the irreducible representation $\Qsl$ of
$\Sigma(n)$ appears in the representation $\Delta^\fvec$.  This is
given by the following integral:
\begin{equation}
  \label{eq:char integral 1}
\nu(\fvec, s\lambda) =  \int_{\Sigma(n)}
d\mu_{\Sigma(n)}\, {{X^\slm}^*}(\Ubar,\sigma) X^\fvec(\Ubar,\sigma).
\end{equation}
Here $X^\slm$ and $X^\fvec$ denote the characters of the
$\Sigma(n)$-representations $\Qsl$ and $\Delta^\fvec$ respectively, and
$d\mu_{\Sigma(n)}$ denotes the normalised Haar measure on $\Sigma(n)$.
Before evaluating the integral \eqref{eq:char integral 1} in Section
\ref{sec:evaluation of integral}, 
we first introduce some background material and notation.

\subsection{Preliminaries}\label{sec:prelim}

\subsubsection{Character formula for $U(k)$}\label{sec:char formula
  for U(k)}

Irreducible representations of the $k$-dimensional unitary group
$U(k)$ are labeled by Young tableaux $\alphavec = (\alpha^1, \ldots,
\alpha^k)$ of $k$ rows (some of which may be empty), where $\alpha^1
\ge \cdots \ge \alpha^k \ge 0$ specify the number of boxes in each row.  Let
\begin{equation}
  \label{eq:|alpha|}
  |\alphavec| = \alpha^1 + \ldots + \alpha^k
\end{equation}
denote the number of boxes in the tableau.  Denote the eigenvalues
of matrices in $U(k)$  by $(\exp \ui \xi_1,\ldots, \exp \ui
\xi_k)$, and their eigenphases by $\xivec = (\xi_1,\ldots,\xi_k)$.
The characters $K_k^\alphavec$ 
of the irreducible representations are functions of $\xivec$, and are
given by the Weyl character formula,
\begin{equation}
  \label{eq:weylformula}
  K^\alphavec_k(\xivec) = 
\frac{
  \left| \begin{array}{cccc}
  \ue^{\ui(\alpha^1+ k-1)\xi_1} & \ue^{\ui(\alpha^1+ k-1)\xi_2} & \cdots & 
  \ue^{\ui(\alpha^1+ k-1)\xi_k}\\
  \ue^{\ui(\alpha^2+ k-2)\xi_1} & \ue^{\ui(\alpha^2+ k-2)\xi_2} & \cdots &
  \ue^{\ui(\alpha^2+ k-2)\xi_k}\\
  \vdots & \vdots & & \vdots \\
  \ue^{\ui \alpha^k\xi_1} & \ue^{\ui\alpha^k \xi_2} & \cdots &
  \ue^{\ui \alpha^k \xi_k}\\
  \end{array} \right| }
  {\left| \begin{array}{cccc}
  \ue^{\ui(k-1)\xi_1} & \ue^{\ui(k-1)\xi_2} & \cdots & 
  \ue^{\ui(k-1)\xi_k}\\
  \ue^{\ui(k-2)\xi_1} & \ue^{\ui(k-2)\xi_2} & \cdots &
  \ue^{\ui(k-2)\xi_k}\\
  \vdots & \vdots & & \vdots \\
  1 & 1 & \cdots & 1\\
  \end{array} \right| } \ .
\end{equation}
Irreducible representations of $SU(k)$ are obtained by restriction.
On $SU(k)$, the representation $\alphavec + r$, which is obtained by
adding $r$ columns of $k$ boxes to $\alphavec$, is equivalent to the
representation $\alphavec$.  $SU(k)$ representations can be uniquely
labeled by Young tableaux of $k-1$ rows (some of which may be empty).

\subsubsection{The Littlewood-Richardson theorem}\label{sec:LR theorem}
Given an irreducible representation $\gammavec$ of $U(k+l)$, its
restriction to the subgroup $U(k) \ttimes U(l)$ is, in general,
reducible, and may be decomposed into a sum of tensor products of 
irreducible representations $\alphavec$ and $\betavec$ of $U(k)$ and $U(l)$, 
respectively. In terms of characters, this decomposition takes the form
\begin{equation}
  \label{eq:Littlewood-Richardson}
  K_{k+l}^\gammavec(\xivec,\etavec) = \sum_{\alphavec,\betavec}
 Y^{\gammavec}_{\boldsymbol{\alpha} \boldsymbol{\beta}} \,
 K_k^{\alphavec}(\xivec)
  K_l^{\betavec}(\etavec),
\end{equation}
where $\xivec$ and $\etavec$ denote the eigenphases of elements of
$U(k)$ and $U(l)$, respectively.  The coefficients
$Y^{\gammavec}_{\boldsymbol{\alpha} \boldsymbol{\beta}}$ in the
decomposition are given by the Littlewood-Richardson theorem,
according to which $Y^{\gammavec}_{\boldsymbol{\alpha}
\boldsymbol{\beta}}$ is the number of times that 
the tableau $\gammavec$ 
can be constructed from $\boldsymbol{\alpha}$ and
$\boldsymbol{\beta}$ by the following procedure:
Boxes from the first row of $\boldsymbol{\beta}$
are added to $\boldsymbol{\alpha}$ so as to produce a new tableau,
with the condition that no two boxes are placed in the same column of
the new tableau.  This is repeated with the second row of
$\boldsymbol{\beta}$, with the additional condition that, on counting
added boxes in the new tableau column-wise from right to left, and
row-wise from top to bottom, the number of added boxes from the first
row of $\betavec$ must always be greater than or equal to the number
of added boxes from the second row.  The procedure is continued for
the other rows until all boxes from $\boldsymbol{\beta}$ have been
added to $\boldsymbol{\alpha}$. It is evident
that  $\gammavec$ can be constructed in this way
only if the number of boxes in $\gammavec$ equals the number
of boxes in $\alphavec$ and $\betavec$ together; that is,
$Y^{\gammavec}_{\boldsymbol{\alpha} \boldsymbol{\beta}}$ vanishes
unless $|\gammavec| = |\alphavec| + |\betavec|$.

Eq.~\eqref{eq:Littlewood-Richardson} generalises to the decomposition of
irreducible representations $\gammavec$ of $U(k_1 + \cdots +
k_c)$ restricted to the subgroup $U(k_1)\ttimes\cdots\ttimes U(k_n)$, as follows:
\begin{equation}
  \label{eq:Littlewood-Richardson multi}
  K_{k_1+\cdots + k_c}^\gammavec(\xivec_1,\cdots,\xivec_c) = 
\sum_{\alphavec_1,\cdots,\alphavec_c}
 Y^{\gammavec}_{\alphavec_1,\ldots,\alphavec_c}
\prod_{b =1}^c  K_{k_b}^{\alphavec_b}(\xivec_b).
\end{equation}
Here the $\alphavec_b$'s are tableaux labeling irreducible
representations of $U(k_b)$, and the $\xivec_b$ denote the eigenphases
of elements of $U(k_b)$.  The $c$-fold coefficients
$Y^{\gammavec}_{\alphavec_1,\ldots,\alphavec_c}$ may be obtained from
the the two-fold coefficients $Y^{\gammavec}_{\boldsymbol{\alpha}
  \boldsymbol{\beta}}$ by performing the $c$-fold decomposition inductively.

With \eqref{eq:Littlewood-Richardson multi}, the $(k_1 + \cdots +
k_c)$-fold determinants in the Weyl character formula
\eqref{eq:weylformula} are reduced to sums of products of ratios of
smaller, $k_b$-fold determinants.  However, this simplification comes
at a price; the Littlewood-Richardson coefficients are not easily
calculated, and closed-form expressions for them are not known.

The original statement of the Littlewood-Richardson theorem appears in
\cite{paper:littlewoodrichardson}.  A modern version with proof may be
found in Macdonald \cite{book:macdonald}.  The application of the
Littlewood-Richardson theorem to the unitary groups 
is discussed by Hagen and MacFarlane  \cite{paper:hagenmacfarlane}
and Itzykson and Nauenberg  \cite{paper:itzyksonnauenberg}.
A more detailed discussion of the rules for multiplying Young tableau
can be found in Hamermesh \cite{book:hamermesh}.

\subsubsection{Characters for $U(2)$ and $SU(2)$}\label{sec:U(2) and SU(2)}

We will need some results and notation particular to the groups $SU(2)$ and 
$U(2)$.
Irreducible characters of $SU(2)$ are denoted by $\chi_{SU(2)}^s(\psi)$,
where $s$ is the spin and $\ue^{\pm \ui \psi}$ denotes the eigenvalues of
elements of $SU(2)$, and are given by
\begin{equation}
  \label{eq:SU(2) characters}
  \chi_{SU(2)}^s(\psi) = \frac{\sin((2s+1)\psi)}{\sin(\psi)}.
\end{equation}
Irreducible representations of $U(2)$ are labeled by tableau
$\alphavec = (\alpha^1,\alpha^2)$ of two rows.  The $U(2)$-characters
$K^\alphavec_2(\xi_1,\xi_2)$ are related to the $SU(2)$-characters 
$\chi_{SU(2)}^s(\psi)$ by
\begin{equation}
  \label{eq:U(2) and SU(2)}
  K_2^\alphavec(+\psi +  \theta, -\psi + \theta) = \ue^{\ui|\alphavec|
  \theta} \chi^{S(\alphavec)}_{SU(2)}(\psi),
\end{equation}
where 
\begin{equation}
  \label{eq:S(alpha)}
  S(\alphavec) = (\alpha^1 - \alpha^2)/2
\end{equation}
denotes the value of spin associated with the $U(2)$-representation 
$\alphavec$.
%Eqs.~\eqref{eq:SU(2)
%characters} and \eqref{eq:U(2) and SU(2)} may be verified from the
%Weyl character formula \eqref{eq:weylformula}.

The Clebsch-Gordan coefficients $C(s_1,s_2,s_3)$ for $SU(2)$ are defined by
\begin{equation}
  \label{eq:clebsh}
  C(s_1, s_2, s_3) = \frac{1}{\pi} \int_0^{2\pi}d\psi \sin^2(\psi) 
\chi^{s_1}_{SU(2)}(\psi) \chi^{s_2}_{SU(2)}(\psi) \chi^{s_3}_{SU(2)}(\psi)
\end{equation}
(note that $\sin^2(\psi/2)$ is the Haar measure with respect to classes
of $SU(2)$), and give the multiplicity of the trivial representation
in the decomposition of the tensor product of 
three $SU(2)$-representations with spins $s_1$, $s_2$ and $s_3$.  
It is an elementary result
that $C(s_1,s_2,s_3)$ equals one if $|s_1 - s_2| \le s_3 \le s_1 +
s_2$, and is zero otherwise.  We define the $r$-fold Clebsch-Gordan
coefficients by
\begin{equation}
  \label{eq:clebsh 2}
  C(s_1, \ldots, s_r) = \frac{1}{\pi} \int_0^{2\pi}d\psi \sin^2(\psi) 
\chi^{s_1}_{SU(2)}(\psi) \cdots  \chi^{s_r}_{SU(2)}(\psi).
\end{equation}
These are given inductively by
\begin{equation}
  \label{eq:clebsch 3}
  C(s_1,\ldots,s_r,s_{r+1}) = \sum_s C(s_1,\ldots,s_{r-1},s) C(s,s_r,s_{r+1}).
\end{equation}

\subsubsection{Cycle decomposition of permutations}\label{sec:cycles}

The following notations will be used for permutations.
Let $\sigma \in S_n$. Denote the factorisation of $\sigma$ into disjoint cycles
by
\begin{equation}
  \label{eq:disjoint cycles}
  \sigma = \sigmahat_1\ldots \sigmahat_{c(\sigma)},
\end{equation}
where $c(\sigma)$ denotes the number of cycles in the factorisation.
Denote the length of a cycle in the decomposition, say $\sigmahat_b$,
by $|\sigmahat_b|$.  

Let $\sigma \in S_n$ and $U = (u_1,\ldots,u_n) \in SU(2)^n$.  For
each cycle $\sigmahat_b$ in the factorisation of
$\sigma$, let $\uhat_b$ denote the product of the corresponding
components of $U$, taken in the reverse order.  That is, if
$\sigmahat_b = (jk\cdots l)$, then
\begin{equation}
  \label{eq:uhat_b}
  \uhat_b = u_l \cdots u_k u_j.
\end{equation}
Similarly, given an $n$-tuple of phases,
$\Theta = (\theta_1,\ldots, \theta_n)$, let
\begin{equation}
  \label{eq:thetahat_b}
  \thetahat_b = \theta_l +  \cdots +  \theta_k +  \theta_j.
\end{equation}
Clearly
\begin{equation}
\label{eq:theta sum}
\theta_1 + \cdots + \theta_n = \thetahat_1 + \cdots + \thetahat_{c(\sigma)}.
\end{equation}

\subsection{Evaluation of the integral}\label{sec:evaluation of integral}

To evaluate the character integral \eqref{eq:char integral 1}, 
it will be convenient to regard $X^\slm$ and
$X^\fvec$ as characters on $SU(2)^n \rtimes \Sigma(n)$, ie as
functions of $U$ and $\sigma$ rather than $\Ubar$ and $\sigma$.  Then
\begin{equation}
  \label{eq:char integral 2}
\nu(\fvec, s\lambda) =  \frac{1}{n!}\sum_{\sigma\in S_n} \int_{SU(2)^n}
du_1 \cdots du_n\, 
{X^\slm(U,\sigma)}^* X^\fvec(U,\sigma),
\end{equation}
where $du_j$ denotes the normalised Haar measure on $SU(2)$.

The character $X^\slm(U,\sigma)$ may be evaluated as follows. 
From \eqref{eq:Q(U)} and
\eqref{eq:Q(sigma)}, 
\begin{eqnarray}
  \label{eq:chi^slm_1}
  X^\slm(U,\sigma) & = & 
\sum_{a = 1}^{d_\lambda} \sum_M 
\bra{M,a}\Qsl(U,\sigma)\ket{M,a}\nonumber\\
%&=&  \sum_{a = 1}^{d_\lambda} \Lambda_{a,a}(\sigma) \sum_M 
%\bra{M}Q^\slm(U,\sigma)\ket{\sigma^{-1}\cdot M}
%\nonumber\\
&=&   \left(\sum_{a = 1}^{d_\lambda} \Lambda_{a,a}(\sigma)\right)
\left(\sum_M 
D^s_{m_{\sigma(1)},m_1}(u_1)\cdots
D^s_{m_{\sigma(n)},m_n}(u_n)\right).\nonumber\\
%&=&  \chi^\lambda_{S_n}(\sigma)\sum_M 
%D^s_{m_{\sigma(1)},m_1}(u_1)\cdots
%D^s_{m_{\sigma(n)},m_n}(u_n) \ ,\nonumber\\
\end{eqnarray}
The sum over $a$  yields $\chi^\lambda_{S_n}(\sigma)$, the character of the 
$S_n$-representation $\Lambda^\lambda$.  The sum over $M$
factorises into a product over the disjoint cycles $\sigmahat_b$ of
$\sigma$, and yields 
\begin{equation}
  \label{eq:traces of D's}
  \Tr D^s(\uhat_1) \cdots \Tr D^s(\uhat_{c(\sigma)}),
\end{equation}
where $\uhat_b \in SU(2)$ is given by \eqref{eq:uhat_b}.  Let
$\ue^{ \pm \ui \xihat_b}$ denote the eigenvalues of $\uhat_b$.  Then
\begin{equation}
  \label{eq:eq:chi slm 3}
   X^\slm(U,\sigma) = 
\chi^\lambda_{S_n}(\sigma)  \prod_{b = 1}^{c(\sigma)}
\chi^s_{SU(2)}(\xihat_b) \ .
\end{equation}
We note that $ X^\slm(U,\sigma)$ is real, since the characters of
$S_n$ and $SU(2)$ are real.

The character $X^\fvec(U,\sigma)$ in \eqref{eq:char integral 2}
may be expressed as
\begin{equation}
  \label{eq:chi fvec}
  X^\fvec(U,\sigma) = \Tr \left(\Gamma^\fvec(x(U,\sigma,\Phi)) P^0\right).
\end{equation}
Here the trace is taken over the carrier space $\Vcal$
of the representation $\Gamma^\fvec$, $\Phi$ denotes phases
satisfying $\ue^{\ui\phi_1}\cdots \ue^{\ui\phi_n} = \sgn(\sigma)$, and
$P^0$ denotes the hermitian projection onto the subspace
$\Vcal^0$ given by
\begin{equation}
  \label{eq:P^0 projection}
  P^0 = \frac{1}{(2\pi)^{n-1}}\int  d^{n-1}\Theta'\,
\Gamma^\fvec(t(\Theta')),
\end{equation}
where the $\Theta'$-integral is taken over $0 \le \theta'_j \le 2\pi$
subject to the condition that 
$\ue^{\ui\theta'_1}\cdots \ue^{\ui\theta'_n} = 1$.
Substituting \eqref{eq:P^0 projection} into \eqref{eq:chi fvec} we get that
\begin{eqnarray}
  \label{eq:chi fvec 2}
  X^\fvec(U,\sigma) &=& \frac{1}{(2\pi)^{n-1}}\int d^{n-1}\Theta'\,
 \Tr \Gamma^\fvec(x(U,\sigma,\Theta' +
  \Phi))\nonumber\\ 
&=& \frac{1}{(2\pi)^{n-1}}\int d^{n-1}\Theta\,
 \Tr \Gamma^\fvec(x(U,\sigma,\Theta)),
\end{eqnarray}
where the integral over $\Theta = (\theta_1,\ldots,\theta_n)$ in the
last expression is restricted to $\ue^{\ui\theta_1}\cdots
\ue^{\ui\theta_n} = \sgn(\sigma)$.  It is convenient to incorporate
this restriction using the identity
\begin{equation}
  \label{eq:Poisson}
\frac{1}{(2\pi)^{n-1}} \int d^{n-1}\Theta = 
\sum_{q=-\infty}^\infty (\sgn \sigma)^q
\frac{1}{(2\pi)^n}
\int \ue^{\ui q(\theta_1 + \cdots + \theta_n)}
\ .
\end{equation}
We note that as  $\Theta$ on the right-hand side of \eqref{eq:Poisson} is 
unconstrained, $x(U,\sigma,\Theta)$ is, in
general, an element of $U(2n)$ rather than $SU(2n)$.  Let $\muvec$
denote the eigenphases of $x(U,\sigma,\Theta)$.  Then 
$ \Tr \Gamma^\fvec(x(U,\sigma,\Theta)) = K^\fvec_{2n}(\muvec)$, and
\eqref{eq:chi fvec 2} becomes
\begin{equation}
  \label{eq:chi fvec 3}
  X^\fvec(U,\sigma) = \sum_{q=-\infty}^\infty (\sgn \sigma)^q
  \frac{1}{(2\pi)^n}\int_0^{2\pi}d^n\Theta \ue^{-\ui q 
(\theta_1 + \cdots \theta_n)} 
K^\fvec_{2n}(\muvec).
\end{equation}

To determine the eigenphases $\muvec$ of $x(U,\sigma,\Theta)$, it is
convenient to represent vectors in $\Cc^{2n}$ as linear combinations of
terms $\ket{v_j}\otimes\ket{j}$, where $\ket{v_j}\in \Cc^2$ and
$\ket{j}$ is an orthonormal basis for $\Cc^n$.  From 
 \eqref{eq:y(sigma,phi)} and \eqref{eq:param x}, 
the action of $x(U,\sigma,\Theta)$ is
then given by
\begin{equation}
  \label{eq:x action}
  x(U,\sigma,\Theta) \sum_{j=1}^n  \ket{\xi_j}\otimes\ket{j} =
 \sum_{j=1}^n \ue^{\ui \theta_j} (u_{\sigma(j)}
 \ket{\xi_j})\otimes\ket{\sigma(j)}\ .
\end{equation}
Let $\sigmahat_b = (jk\cdots l)$ be a cycle in $\sigma$, and, as above, let 
$\ue^{\pm \ui \xihat_b}$ denote the eigenvalues of $\uhat_b$.  Let
$\ket{\pm w_b} \in \Cc^2$ denote the associated eigenvectors of $\uhat_b$.  
It is readily
verified that $\ket{\pm w_b}\otimes \ket{l}$ are eigenvectors of
$x^{m}(U,\sigma,\Theta)$, with eigenvalues $\ue^{\ui m(\pm \xihat_b +
\thetahat_b)/|\sigmahat_b|}$, if and only if $m$ is a multiple of
$|\sigmahat_b|$.

In general, if $\ket{v}$ is an eigenvector of some positive integer
power $|\sigmahat_b|$ of a matrix $M$, with $\rho$ the associated
eigenvalue of $M^{|\sigmahat_b|}$, and if $\ket{v}$ is not an
eigenvector of any smaller positive power of $M$, then $M$ has
eigenvalues $\ue^{2\pi \ui p / |\sigmahat_b|}\rho^{1/|\sigmahat_b|}$,
where $p = 1,\ldots,|\sigmahat_b|$.  From these considerations we may
deduce that to each cycle $\sigmahat_b$ in $\sigma$ are associated
$2|\sigmahat_b|$ eigenphases of $x(U,\sigma,\Theta)$, denoted by
$\etavec_b = (\etavec_{b,1},\ldots,\etavec_{b,|\sigmahat_b|})$ and given 
explicitly by
\begin{equation}
  \label{eq:etavec_b,p}
  \etavec_{b,p} = \left(\frac{ +\xihat_b + \thetahat_b + 2\pi p}{
|\sigmahat_b|},
\frac{ -\xihat_b+\thetahat_b + 2\pi p}{|\sigmahat_b|}\right), \quad  
p = 1,\ldots, |\sigmahat_b|.
\end{equation}
The full set of eigenvalues of
$x(U,\sigma,\Theta)$ is 
\begin{equation}
  \label{eq:muvec formula}
  \muvec = (\etavec_1,\ldots,\etavec_{c(\sigma)})
\end{equation}

From \eqref{eq:muvec formula} it is apparent that $x(U,\sigma,\Theta) \in
U(2n)$ is unitarily equivalent to the element of
$U(2|\sigma_1|)\ttimes\cdots \ttimes U(2|\sigma_{c(\sigma)}|)$
with eigenphases $\etavec_1, \ldots, \etavec_{c(\sigma)}$ for the factors.
From the Littlewood-Richardson formula
\eqref{eq:Littlewood-Richardson multi}, the character
$K_{2n}^\fvec(\muvec)$ 
in \eqref{eq:chi fvec 3} is given
by
\begin{eqnarray}
  \label{eq:chifvec LR}
  K_{2n}^\fvec(\muvec) &=&
  K_{2n}^\fvec(\etavec_1,\ldots,\etavec_{c(\sigma)})\nonumber\\
&=& \sum_{\betavec_1,\cdots,\betavec_{c(\sigma)}}
 Y^{\fvec}_{\betavec_1,\ldots,\betavec_{c(\sigma)}}
\prod_{b=1}^{c(\sigma)}  K_{|\sigmahat_b|}^{\betavec_b}(\etavec_b),
\end{eqnarray}
where the $\betavec_b$'s are tableaux labeling representations of
$U(2|\sigmahat_b|)$.  As noted in Section~\ref{sec:LR theorem}, 
the sum in \eqref{eq:chifvec LR} may restricted to those $\betavec_b$
satisfying
\begin{equation}
  \label{eq:beta sum}
  \sum_{b = 1}^{c(\sigma)} |\betavec_b| = |\fvec|.
\end{equation}
From \eqref{eq:etavec_b,p}, the characters
$K_{|\sigmahat_b|}^{\betavec_b}(\etavec_b)$ can themselves be
expressed as a product of $U(2)$-characters by applying
the Littlewood-Richardson theorem once more, as follows:
\begin{eqnarray}
  \label{eq:chibetavec LR}
K_{|\sigmahat_b|}^{\betavec_b}(\etavec_b) &=&
K_{|\sigmahat_b|}^{\betavec_b}(\etavec_{b,1},\ldots,\etavec_{b,|\sigmahat_b|})
\nonumber\\
&=& \sum_{\alphavec_1,\cdots,\alphavec_{|\sigmahat_b|}}
 Y^{\betavec_b}_{\alphavec_1,\ldots,\alphavec_{|\sigmahat_b|}}
\prod_{p=1}^{|\sigmahat_b|}  K_2^{\alphavec_p}(\etavec_{b,p}),
\end{eqnarray}
Here the $\alphavec_p$'s are tableaux labeling representations of
$U(2)$, and the sum in \eqref{eq:chibetavec LR}
may be restricted to those $\alphavec_p$ satisfying
\begin{equation}
  \label{eq:alpha sum}
  \sum_{p = 1}^{|\sigmahat_b|} |\alphavec_p| = |\betavec_b|.
\end{equation}
Finally, the $U(2)$-characters are given explicitly (cf
\eqref{eq:U(2) and SU(2)} and \eqref{eq:etavec_b,p}) by
\begin{equation}
  \label{eq:chi alphavec}
 K_2^{\alphavec_p}(\etavec_{b,p}) = \ue^{\ui|\alphavec_p|(2\pi p + \thetahat_b)
/|\sigmahat_b|} \chi_{SU(2)}^{S(\alphavec_p)}(\xihat_b/|\sigmahat_b|).
\end{equation}

To proceed, we substitute the expression \eqref{eq:eq:chi slm 3} for
$X^\slm(U,\sigma)$ and the expressions \eqref{eq:chi fvec 3} and
\eqref{eq:chifvec LR} -- \eqref{eq:chi alphavec} for
$X^\fvec(U,\sigma)$ into the integral \eqref{eq:char integral 2}.  The
integration over $SU(2)^n$ can be arranged so that $\uhat_1, \ldots,
\uhat_{c(\sigma)} \in SU(2)$ are amongst the integration variables.
Since the integrand depends only on the $\uhat_b$'s, any remaining
$SU(2)$-integrals are trivially evaluated.  Moreover, since the
integrand depends only on the eigenphases $\xihat_b$, we can make the
replacement
\begin{equation}
  \label{eq:SU(2) measure to class measure}
  \int_{SU(2)}d\uhat_b \rightarrow \frac{1}{\pi}
\int_0^{2\pi} d\xihat_b \sin^2(\xihat_b).
\end{equation}
Similarly, the $\Theta$-integral in \eqref{eq:char integral 2} can be
arranged so that $\thetahat_1, \ldots, \thetahat_{c(\sigma)}$ are
amongst the variables of integration; in view of \eqref{eq:theta sum},
the integrand depends only on the $\thetahat_b$'s, so the integrals
over any remaining components of $\Theta$ are trivially evaluated.
Finally, as the terms in the $S_n$-sum in \eqref{eq:char integral 2}
depend only the conjugacy class of $\sigma$ and 
not on $\sigma$ itself, we may make the replacement
\begin{equation}
\label{eq: S_n sum over classes}
\sum_{\sigma \in S_n} \rightarrow \sum_{[\sigma]\in S_n} \Omega_{[\sigma]},
\end{equation}
where $[\sigma]$ denotes the conjugacy class of $\sigma$ and
$\Omega_{[\sigma]}$ denotes the number of elements in $[\sigma]$
($\Omega_{[\sigma]}$ may be explicitly expressed in terms of the cycle
lengths $|\sigmahat_1|, \ldots, |\sigmahat_{c(\sigma)}|$).  In this
way, Eq.~\eqref{eq:char integral 2} for the multiplicities may be expressed 
\begin{equation}
  \label{eq:nu(f,sl) 2} 
  \nu(\fvec,\slm) =  
  \frac{1}{n!}
  \sum_{[\sigma]\in S_n} \Omega_{[\sigma]}
  \chi^\lambda_{S_n}(\sigma)\sum_{q = -\infty}^{\infty} (\sgn
  \sigma)^q \sum_{\betavec_1,\cdots,\betavec_{c(\sigma)} }
  Y^{\fvec}_{\betavec_1,\ldots,\betavec_{c(\sigma)}}\prod_{b=1}^{c(\sigma)}
I_b\, J_b \ .
\end{equation}
The factor $I_b$, which contains the integral over $\theta_b$, is given by
\begin{equation}
\label{eq:I_b}
I_b = \frac{1}{2\pi}
 \int_0^{2\pi}d\thetahat_b \ue^{\ui(|\betavec_b|/ |\sigmahat_b| -
 q)\thetahat_b},
\end{equation}
where we have used \eqref{eq:alpha sum}.
The factor $J_b$, which contains the integral over $\xihat_b$ and the
sum over the $U(2)$-tableaux $\alphavec_p$,  is given by
\begin{eqnarray}
\label{eq:J_b}
\lefteqn{J_b  =
\sum_{\alphavec_1,\cdots,\alphavec_{|\sigmahat_b|}}
Y^{\betavec_b}_{\alphavec_1,\ldots,\alphavec_{|\sigmahat_b|}}
\ue^{2\pi \ui \left(\sum_{p=1}^{|\sigmahat_b|}p |\alphavec_p|\right)
/|\sigmahat_b|} \times \frac{1}{\pi}\times } \nonumber\\
&&  
\int_0^{2\pi} d\xihat_b 
\sin^2(\xihat_b) \chi^s_{SU(2)}(\xihat_b)
\chi^{S(\alphavec_1)}_{SU(2)}
(\xihat_b/|\sigmahat_b|)
%\left(\frac{\xihat_b}{|\sigmahat_b|}\right)
\cdots 
\chi^{S(\alphavec_{|\sigmahat_b|})}_{SU(2)}
%\left(\frac{\xihat_b}{|\sigmahat_b|}\right)
(\xihat_b/|\sigmahat_b|) \ .
\end{eqnarray}
%(HAD TROUBLE TEX-ING THE PRECEDING EQUATION - AM STILL NOT SATISFIED)
%
%
%\begin{multiline}\label{eq:J_b}
%J_b  =
%\sum_{\alphavec_1,\cdots,\alphavec_{|\sigmahat_b|}}
%Y^{\betavec_b}_{\alphavec_1,\ldots,\alphavec_{|\sigmahat_b|}}
%\ue^{2\pi \ui \left(\sum_{p=1}^{|\sigmahat_b|}p |\alphavec_p|\right)
%/|\sigmahat_b|} \times} \\
%   \times 
%\frac{1}{\pi} \int_0^{2\pi} d\xihat_b 
%\sin^2(\xihat_b/2) \chi^s_{SU(2)}(\xihat_b)
%\chi^{S(\alphavec_1)}_{SU(2)}
%(\xihat_b/|\sigmahat_b|)
%\left(\frac{\xihat_b}{|\sigmahat_b|}\right)
%\cdots 
%\chi^{S(\alphavec_{|\sigmahat_b|)}}_{SU(2)}(\xihat_b/|\sigmahat_b|)\reqno.
%\end{multiline}

%\left(\frac{\xihat_b}{|\sigmahat_b|}\right)
%\frac{1}{(2\pi)^2}  \sum_{q=-\infty}^\infty (\sgn \sigma)^q
%\int_0^{2\pi} d\thetahat_\beta \ue^{\ui(\beta^b/|\sigmahat_b| -
% q)\thetahat_b)} 
%\int_0^{2\pi} d\xihat_b \sin^2(\xihat_b/2)
%\chi^2_{SU(2)}(\xihat_b) \prod_{l=1}^{|\sigmahat_b|} 
%\ue^{2\pi i l |\alpha_l|/|\sigmahat_b|}
% \chi^{S(\alpha^l)}_{SU(2)}(\xihat_b/|\sigmahat_b|).
%\end{equation}

The $\thetahat_b$-integral in \eqref{eq:I_b} is trivial, and vanishes unless
\begin{equation}
\label{eq:I integral result}
 q |\sigmahat_b| = |\betavec_b|.
\end{equation}
Summing over $b$ in \eqref{eq:I integral result} and
using \eqref{eq:beta sum}, we get that
\begin{equation}
  \label{eq:q and beta^b}
  q = \frac{|\fvec|}{n}.
\end{equation}
Since $q$ is an integer, it follows that at least one of the $I_b$'s
must vanish (and, therefore, $\nu(\fvec,s\lambda)$ must vanish) unless
$n$ divides $|\fvec|$.  Assuming this to be so, the sum over $q$ in
\eqref{eq:nu(f,sl) 2} collapses to $q = |\fvec|/n$.

We consider next the expression for $J_b$.  Since the integrand is
$2\pi$-periodic in $\xihat_b$, we can make the replacement
\begin{equation}
  \label{eq:replace}
  \int_0^{2\pi} d\xihat_b \rightarrow 
%\frac{1}{|\sigmahat_b|}
%  \int_0^{2\pi|
%\sigmahat_b|} d\xihat_b \rightarrow 
  \int_0^{2\pi}
d\psihat_b,
\end{equation}
where
$\psihat_b = \xihat_b/|\sigmahat_b|$. Using the identity
\begin{equation}
  \label{eq:su(2) char identity}
  \sin^2(r\psihat_b)\chi^s_{SU(2)}(r\psihat_b)  = 
\sin^2(\psihat_b)\chi^{rs+(r-1)/2}_{SU(2)}(\psihat_b)
\chi^{(r-1)/2}_{SU(2)}(\psihat_b) \  , 
\end{equation}
which follows from the definition \eqref{eq:SU(2) characters},  the
resulting integral over $\psihat_b$ is of the form \eqref{eq:clebsh
2}, and yields a $(|\sigmahat_b| + 2)$-fold Clebsch-Gordan
coefficient.  We obtain
\begin{eqnarray}
  \label{eq:J_b final}
\lefteqn{ J_b =
\sum_{\alphavec_1,\cdots,\alphavec_{|\sigmahat_b|}}
Y^{\betavec_b}_{\alphavec_1,\ldots,\alphavec_{|\sigmahat_b|}}
\ue^{2\pi \ui \left(\sum_{p=1}^{|\sigmahat_b|}p |\alphavec_p|\right)
/|\sigmahat_b|} \times} \nonumber\\
& & \times \,
C\left(|\sigmahat_b|s +  \half(|\sigmahat_b| -1), 
\half(|\sigmahat_b| -1),
S(\alphavec_1),\ldots,
  S(\alphavec_{|\sigmahat_b|})\right) \ .
\end{eqnarray}
From \eqref{eq:nu(f,sl) 2}, \eqref{eq:q and beta^b} and \eqref{eq:J_b
final}, we get our main result for the multiplicities,
\begin{equation}
\label{eq:main 1}
\nu(\fvec,\slm) = \frac{1}{n!}
  \sum_{[\sigma]\in S_n} \Omega_{[\sigma]} \chi^\lambda_{S_n}(\sigma)
 (\sgn
  \sigma)^{|\fvec|/n} A_{[\sigma]},
\end{equation}
where
\begin{equation}
\label{eq:main2}
\begin{split} A_{[\sigma]} = & \mathop{\sum_{\betavec_1,\cdots,
\betavec_{c(\sigma)}}}_{|\betavec_b| = |\sigmahat_b|\cdot|\fvec|/n}
Y^{\fvec}_{\betavec_1,\ldots,\betavec_{c(\sigma)}}
\times \prod_{b=1}^{c(\sigma)}\Bigg [
\sum_{\alphavec_1,\cdots,\alphavec_{|\sigmahat_b|}}
Y^{\betavec_b}_{\alphavec_1,\ldots,\alphavec_{|\sigmahat_b|}}
\ue^{2\pi \ui \left(\sum_{p=1}^{|\sigmahat_b|}p |\alphavec_p|\right)
|\sigmahat_b|} \times \\ 
& \times C\left(|\sigmahat_b|s +  \half(|\sigmahat_b| -1), 
\half(|\sigmahat_b| -1),
S(\alphavec_1),\ldots,
  S(\alphavec_{|\sigmahat_b|})\right)\Bigg ] \ .
%C\left(|\sigmahat_b|s + \frac{|\sigmahat_b| -1}{2}, 
%\frac{|\sigmahat_b| -1}{2},
%S(\alphavec_1),\ldots,
%  S(\alphavec_{|\sigmahat_b|})\right)
\end{split}
\end{equation}

For $\sigma = \UI = (1)\cdots (n)$ (ie, the identity element in $S_n$),
the expression \eqref{eq:main2} simplifies considerably.  In this
case, $c(\sigma) = n$ and $|\sigma_b| = 1$, so that each $\betavec_b$
is a $U(2)$-tableaux with $|\betavec_b| = |\fvec|/n$. The sum over
$\alphavec_p$ collapses to  $\alphavec = \betavec_b$, and
the corresponding Clebsch-Gordan coefficient, 
$C(s, 0, S(\betavec_b))$, vanishes unless 
$S(\betavec_b) = s$.  It follows that
the $\betavec_b$ must all coincide with the $U(2)$-tableaux
$\betavec(\fvec,n,s)$ given by
\begin{equation}\label{eq: betavec fvec}
\betavec(\fvec,s) = \left((|\fvec|/2n + s), (|\fvec|/2n - s)
\right) \ .
\end{equation}
We then obtain 
\begin{equation}\label{eq: A_I}
A_{[I]} =
Y^\fvec_{\underbrace{\betavec(\fvec,s),\cdots,\betavec(\fvec,s)}_{\hbox{$n$\
\rm times}}}.
\end{equation}
Some simplification in~\eqref{eq:main2}  also occurs for $n$-cycles
in $S_n$, eg
$\sigma = (12\cdots n)$,
In this case $c(\sigma) = 1$ and
$|\sigmahat| = n$, so that the sum over
$\betavec$ collapses to $\betavec = \fvec$.  We obtain
\begin{eqnarray}\label{eq: n cycle}
\lefteqn{A_{[(12\cdots n)]} = 
\sum_{\alphavec_1,\cdots,\alphavec_n}
Y^{\fvec}_{\alphavec_1,\ldots,\alphavec_n}
\ue^{2\pi \ui \left(\sum_{p=1}^{n}p |\alphavec_p|\right)
/n}\, \times} \nonumber \\
& & 
\times\,C\left(ns +  \half(n -1), 
\half(n -1),
S(\alphavec_1),\ldots,
  S(\alphavec_n)\right) \ .
\end{eqnarray}

The sum $\sum_\lambda \nu(\fvec,s\lambda)$ gives the number of
$n$-spin-$s$ representations, regardless of statistics.  Using the
character relation (see, eg, \cite{book:hamermesh}),
\begin{equation}\label{eq: char sum}
\sum_\lambda \chi^\lambda(\sigma) = 
\left\{
\begin{array}{ll}
n!,& \sigma = \UI,\\
0, & \hbox{\rm otherwise}, 
\end{array}
\right.
\end{equation}
we obtain from \eqref{eq:main 1} and \eqref{eq: A_I} 
a simple expression for these summed multiplicities,
\begin{equation}\label{eq: sum of nus}
\sum_\lambda \nu(\fvec,\slm) = 
Y^\fvec_{\underbrace{\betavec(\fvec,s),\cdots,\betavec(\fvec,s)}_{\hbox{$n$\
\rm times}}}.
\end{equation}

%WORK FROM HERE ***************************************************

\subsection{Examples}

In Section~\ref{sec:spin-stat relations}, we considered the case of
completely symmetric representations, for which $\fvec = (d)$.  There
we showed that $\nu((d),s\lambda)$ vanishes unless $d = 2ns$ and
unless $\lambda$ is the trivial (resp.~alternating) representation
according to whether $s$ is integral (resp.~half-odd-integral).  This
particular case is readily obtained from the general formulas
(\ref{eq:main 1}) and (\ref{eq:main2}).  Instead of doing so, we
sketch below the analogous calculation for the completely
antisymmetric representations of $SU(2n)$.  The result turns out to be
rather different from the symmetric case, in that only $s=\half$
constructions are supported; the completely antisymmetric
representations do not provide a systematic description for all spins.
The $s=\half$ statistics turn out to be bosonic in this case.

Completely antisymmetric representations of $SU(2n)$ correspond to single-columned
tableaux of between 1 and $2n-1$ rows (the $2n$-rowed tableau is
equivalent to the trivial representation).  Denote the $d$-rowed
representation by $\fvec = (1)^d$.
From (\ref{eq:q and beta^b}), $\nu((1)^d,s\lambda)$ vanishes unless $d
= n$ and, from (\ref{eq: betavec fvec}) and (\ref{eq: sum of nus}),
unless $s = \half$.  In this case, the expressions (\ref{eq:main 1})
and (\ref{eq:main2}) simplify considerably.  
The sums over the
$\alphavec_p$'s collapse
to the single term where all the $|\alphavec_p|$'s are equal to one, and
the sums over the 
$\betavec_b$'s collapse to the single term where 
$\betavec_b = (1)^{|\sigmahat_b|}$.
For these terms, the Littlewood-Richardson coefficients and
Clebsch-Gordan coefficients appearing in (\ref{eq:main2}) are all
equal to one.  We get that
\begin{equation}
  \label{eq:A sigma for column}
  A_{[\sigma]} = \prod_{b=1}^{c(\sigma)}
\ue^{2\pi \ui \left(\sum_{p=1}^{|\sigmahat_b|}p \right)
/|\sigmahat_b|} = \prod_{b=1}^{c(\sigma)}(-1)^{|\sigmahat_b| + 1} = 
\prod_{b=1}^{c(\sigma)} \sgn(\sigmahat_b) = \sgn(\sigma).
\end{equation}
Substituting the preceding into (\ref{eq:main 1}), we obtain
\begin{equation}
  \label{eq:nu_a}
  \nu((1)^n,\half\ \lambda) = \frac{1}{n!}
\sum_{[\sigma]\in S_n} \Omega_{[\sigma]} \chi^\lambda_{S_n}(\sigma),
\end{equation}
which vanishes unless $\lambda$ is the trivial
representation of $S_n$.

This result can also be obtained by following the calculation of
Section~\ref{sec:spin-stat relations} and regarding $z =
(z_1,\ldots,z_{2n})$ as Grassmann variables.  Equivalently, this may
be regarded as the $n$-particle version of the
`anti-Schwinger' construction of \cite{paper:berryrobbins2}, wherein
the raising/lower operators of \cite{paper:berryrobbins} are made
to satisfy anticommutation relations. The anticommutation
relations are responsible for the restriction to $s = \half$.

Next, we consider general representations for the case of two
particles, $n =2$.
For simplicity,
we label the even and odd representations of $S_2$ by $\lambda = +$
and $\lambda = -$, respectively, with characters
$\chi_{S_2}^\pm(\sigma) = \pm \sgn(\sigma)$.  As in \eqref{eq: betavec
fvec}, let
\begin{equation}\label{eq: betavec fvec n = 2}
\betavec(\fvec,s) = \left((|\fvec|/4 + s), (|\fvec|/4 - s)
\right). \ 
\end{equation}
From \eqref{eq: betavec fvec n = 2} we can
deduce that the multiplicities $\nu(\fvec,s \pm)$ 
vanish unless $|\fvec|$ is even and
$|\fvec|/4 - s$ is a nonnegative integer.  From
\eqref{eq:main 1}, \eqref{eq: A_I} and \eqref{eq: n cycle}, 
we get
\begin{eqnarray} \label{eq:char integral n=2 res}
\lefteqn{ \nu ( \fvec, s \pm )
  = \frac{1}{2}
  Y_{\betavec(\fvec,s), \betavec(\fvec,s)}^{\mathbf{f}} 
\pm  } \nonumber \\ & & \pm
  \frac{(-1)^{2s}}{2} \sum_{\alphavec_1,\alphavec_{2}} 
  Y^{\fvec}_{\alphavec_1,\alphavec_2} 
 (-1)^{|\alphavec_1|} 
 C(2s+\half, \half, S(\alphavec_1), S(\alphavec_2)) \ .
\end{eqnarray}

The simplest cases are the tableaux $\fvec = (2,0)$ and $\fvec =
(1,1)$ with $|\fvec| = 2$.  Then $s = \half$ is the only permitted
value of the spin, and $\betavec(\fvec,s)$ contains a single box.  The
tableau $(2,0)$ corresponds to a completely symmetric representation.
As discussed in Section~\ref{sec:spin-stat relations}, this yields
fermi statistics for spin-$\half$, so that $\nu(\fvec, \half +) = 0$ and
$\nu(\fvec, \half -) = 1$.
\eqref{eq:char integral n=2 res} is readily evaluated for the tableau $(1,1)$,
and one finds the opposite result, namely bose statistics for
spin-$\half$ (this case is equivalent to the ``anti-Schwinger
construction'' discussed in \cite{paper:berryrobbins2}).

For larger tableaux one typically finds that for each permitted value
of spin, both multiplicities $\nu(\fvec, s +)$ and $\nu(\fvec, s -)$
are nonzero. Constructions based on such tableaux do not determine a
spin-statistics relation.  The smallest tableau where this occurs is
$\fvec = (3,2,1)$ with $s = \half$.  Evaluation of \eqref{eq:char
integral n=2 res} shows there is one fermi and one bose
representation.  In fact, for larger tableaux one typically finds that
$\nu(\fvec, s +)$ and $\nu(\fvec, s -)$ are nearly equal, and there
are arguments to suggest they are either exactly equal, or else differ
by 1, according to whether their sum, given by
$Y_{\betavec(\fvec,s), \betavec(\fvec,s)}^{\mathbf{f}}$, is even or
odd.  It turns out that most of the terms in the sum over tableaux in
\eqref{eq:char integral n=2 res} cancel.  First, since the summand is
symmetric in $\alphavec_1$ and $\alphavec_2$, apart from the sign
factor $(-1)^{|\alphavec_1|}$, only terms for which $|\alphavec_1|$ and
$|\alphavec_2|$ have the same parity contribute.  Amongst these
remaining terms, the sign factor is responsible for additional
cancellations, due to the following fact: If $\alphavec_1 + \epsilon$
is the tableau obtained by adding one box to the first row of
$\alphavec_1$, and $\alphavec_2 - \epsilon$ is the tableau obtained by
removing one box from the first row of $\alphavec_2$, then the rules
for multiplying tableaux imply that
\begin{equation}
  Y^\fvec_{\alphavec_1,\alphavec_2} = 
  Y^\fvec_{\alphavec_1+\epsilon,\alphavec_2-\epsilon}.
\end{equation}
Details may be found in \cite{thesis:harrison}.

To demonstrate the possibility of parastatistics, we consider the simplest 
case
of three particles and the smallest $SU(6)$ tableau, $\fvec = (2,1)$.
From \eqref{eq: betavec fvec}, it follows that $s=\half$ is the only
permitted value of spin, and that $\betavec(\fvec,\half) = (1)$ consists of
a single box.
Let $\lambda = E$ denote the two-dimensional 
representation of $S_3$ (corresponding to the tableau $(2,1)$), with
characters 
\begin{equation}
  \chi_{S_3}^{E}(\UI)= 2,   \quad \chi_{S_3}^{E}((12)) = 0, 
  \quad \chi_{S_3}^{E}((123)) = -1,
\end{equation}
and classes 
\begin{equation}\label{eq: classes}
\Omega_{[\UI]} = 1, \quad \Omega_{[(12)]} = 3, 
\quad \Omega_{[(123)]} = 2
\end{equation}
(see, eg \cite{book:hamermesh}).
From 
\eqref{eq:main 1}, \eqref{eq: A_I} and \eqref{eq: n cycle}, 
we get
\begin{eqnarray} \label{eq:char integral n=3 res}
\lefteqn{\nu((2,1), \half E)
   = \frac{1}{6}
  Y_{(1), (1), (1)}^{(2,1)} }\nonumber\\
& &
  - \frac{1}{6} \times 2 \sum_{\alphavec_1,\alphavec_{2}, \alphavec_{3}} 
  Y^{(2,1)}_{\alphavec_1,\alphavec_{2},\alphavec_3} 
\ue^{\ui \frac{2\pi}{3}(|\alphavec_1| + 2|\alphavec_2|)}
%  \times  \nonumber \\ & & 
  C(\fhalf, 1, S(\alphavec_1), S(\alphavec_2), S(\alphavec_3)) \ .
\end{eqnarray}
There are four sets of $U(2)$-tableaux $\alphavec_1$, $\alphavec_2$
and $\alphavec_3$ which may be multiplied to obtain the $SU(6)$
tableau $(2,1)$ (including cases where one or more of the $\alphavec_j$ 
are empty, which we denote by $\alphavec_j = 0$). 
For each combination we determine the
Littlewood-Richardson coefficient
$Y^{(2,1)}_{\alphavec_1,\alphavec_{2},\alphavec_3}$, the Clebsch-Gordan
coefficient $C(\fhalf, 1, S(\alphavec_1), S(\alphavec_2),
S(\alphavec_3))$, and the phase factor $\ue^{\ui
\frac{2\pi}{3}(|\alphavec_1| + 2|\alphavec_2|)}$.  The results are
summarised in the table below.
\begin{equation}\label{eq: data}
\begin{array}{ccclccc}
\alphavec_1&\alphavec_2&\alphavec_3&\vline&Y&C&\ue^{\ui \frac{2\pi}{3}}\\
\hline
(2,1)&0&0  &\vline  &1&0&1\\
(1,1)&(1)&0&\vline  &1&0&\ue^{\ui \frac{2\pi}{3}}\\
(2)&(1)&0&  \vline  &1&1&\ue^{\ui \frac{2\pi}{3}}\\
(1)&(1)&(1)&\vline  &2&1&1
\end{array}
\end{equation}
Substituting \eqref{eq: data} into \eqref{eq:char integral n=3 res},
we obtain
\begin{equation} \label{eq:char integral n=3 res 2}
  \nu((2,1), \half E) = 1.
\end{equation}
That is, a $3$-spin-$1/2$ 
bundle with parastatistics 
may be constructed from the representation $\Gamma^{(2,1)}$ of $SU(6)$.

\section{Discussion}\label{sec:discussion}

We have reformulated the quantum
kinematics of BR for indistinguishable spinning particles in terms of
vector bundles over $n$-particle configuration space.
Within this geometrical framework, our main results concern a
repre\-sentation-theoretic generalisation of the construction in BR.  We
have shown that $n$-spin bundles can be constructed from irreducible
representations $\Gamma^\fvec$ of the group $SU(2n)$.  The
construction makes use of representations $\Delta^\fvec$ of the
spin-statistics group $\Sigma(n)$ associated to $\Gamma^\fvec$, as
well the existence of a continuous, $S_n$-equivariant map from
$SU(n)/T(n)$ to configuration space $C_n$ 
\cite{paper:atiyah1}.

The construction in BR is based on particular representations of
$SU(2n)$, namely the completely symmetric representations. For a given
number, $n$, of indistinguishable particles with spin $s$, there is a
unique completely symmetric representation of $SU(2n)$, namely the
$2ns$-fold symmetric tensor product of $SU(2n)$ with itself, which
leads to a description of the quantum kinematics (ie, which supports
an $n$-spin bundle with spin $s$).  The statistics is necessarily in
accord with the physically correct spin-statistics relation.

Representations of $SU(2n)$ other than the completely symmetric
representations, corresponding to Young tableaux $\fvec$ of more than
one row, typically support multiple values of spin $s$, and for a
given spin may support distinct values of the statistics $\lambda$,
including parastatistics.  The values of spin and statistics supported
by a given representation $\Gamma^\fvec$ are determined by the
multiplicities $\nu(\fvec,\slm)$ of the irreducible representations
of the spin-statistics group, $\Sigma(n)$, in the decomposition of
$\Delta^\fvec$.

Our main calculation is an evaluation of the multiplicities
$\nu(\fvec,\slm)$ using character methods. 
Eqs.~\eqref{eq:main 1}--\eqref{eq:main2} 
give the multiplicities as a finite sum over
characters of the symmetric group $S_n$, the $n$-fold Clebsch-Gordan
coefficients of $SU(2)$, and the Littlewood-Richardson coefficients
for the decomposition of representations of $U(n+m)$ into
representations of $U(n) \times U(m)$.  

Our calculation is related to a more general problem in representation
theory, namely the decomposition of zero-weight representations of the
Weyl group $W$ of a compact, connected Lie group $G$ associated with
an irreducible representation of $G$. It would be interesting to see
if alternative methods could be brought to bear on our calculation,
as well as whether the methods used here might prove useful in other contexts.  Our
construction of $n$-spin-bundles is similarly related to the
construction of flat zero-weight bundles over the coset space $G/T$
(where $T$ is a maximal torus of $G$), whose decomposition into a direct sum
of sub-bundles irreducible under monodromy leads to the decomposition
problem described above.

Concerning the spin-statistics relation, in the first instance our
results are similar to those of 
\cite{paper:berryrobbins}.  Within the
group-theoretical framework considered here, the requisite properties
introduced in Section~\ref{sec:bundle} do not determine a connection
between spin and statistics.  When general representations of $SU(2n)$
are admitted alongside the completely symmetric representations, the
spin-statistics relation is lost.

As argued in the Introduction, a derivation of the spin-statistics
relation from a reformulation of quantum mechanics should be based on
principles whose physical motivation is clear. The role played by the
group $SU(2n)$ in our considerations is not well motivated in this respect.
One could offer as motivation the fact that $SU(2n)$ incorporates both
rotations of $n$ spins (ie, $SU(2)^n$) and the permutations $S_n$, but of
course it is not the only group which does so. 

However, the role played by the completely symmetric representations
deserves further consideration.  They provide, at least as far as we
have discerned, the only systematic means, within the given framework,
of associating a representation to a particular value of spin.  It is
suggestive, too, that the scheme which works treats the spins in a
completely symmetrical way; this would seem appropriate for
indistinguishable particles.  Indeed, characteristic aspects of the
completely symmetric representations may indicate a different approach
to this nonrelativistic treatment of the spin-statistics relation,
which we hope to report on in future.

\bigskip

\noindent{\bf Acknowledgements.}  We thank Michael Berry, Roe Goodman,
Aidan Schofield and David Thouless for helpful discussions.  JMH was
supported by the EPSRC and the European Commission under the Research
Training Network (Mathematical Aspects of Quantum Chaos) no.
HPRN-CT-2000-00103 of the IHP Programme.  JMR acknowledges the
hospitality of the Mathematical Sciences Research Institute while the
manuscript was being completed.

\appendix
\section{General Setting}\label{sec:generalised}

The construction of $n$-spin bundles described in
Section~\ref{sec:su2n} is closely related to the following general
problem (see, eg,
\cite{paper:kostant}).  Let $\Gamma$ be an irreducible unitary 
representation on a
vector space $\Vcal$ of a compact Lie group $G$ with maximal torus
$T$.  $\Vcal$ may be decomposed into weight spaces, $\Vcal^\mu$, labeled
by weights $\mu$ of $T$.  The zero-weight space $\Vcal^0$ carries a
representation $\Delta$ of the Weyl group, $W$, of $G$, which is, in
general, reducible.  One can then ask for the decomposition of this
representation of the Weyl group into its irreducible components.

To each weight space $\Vcal^\mu$ of the representation $\Gamma$ is
associated a hermitian vector bundle $\Ecal^\mu$ over $G/T$ with
Abelian $G$-invariant hermitian connection. The curvature of this
connection depends linearly on $\mu$, and therefore
vanishes on the zero-weight bundle $\Ecal^0$. If $G$ is simply
connected, then $\Ecal^0$ is trivial.  However, the quotient bundle
$\Ebarcal^0 = \Ecal^0/W$ over the quotient space $(G/T)/W$, while locally flat,
may be nontrivial.  For simply connected $G$, the fundamental group of
the quotient space is just the Weyl group $W$, and the monodromy of
the flat connection yields a representation of $W$, which is precisely
the representation $\Delta$ described above.

This setting can be further generalised by regarding $G$ as a subgroup
of a Lie group $F$, and regarding $\Gamma$ as the restriction to $G$
of a representation of $F$. In this case, the natural
structure group for the zero-weight bundle is the generalised Weyl
group $V = M/T$, where $M$ is the $F$-normaliser of $T$.

The construction of Section~\ref{sec:su2n} is an example belonging to
this more general setting, with $F = SU(2n)$, $G = SU(n)$, and $T =
T(n)$.  The Weyl group of $SU(n)$ is just $S_n$, and the generalised
Weyl group $V$ is the spin-statistics group, $\Sigma(n) =
SU(2)^n\rtimes S_n/\Null(n)$.  The $n$-spin bundles are pullbacks, via the
$S_n$-equivariant map $\Xi: C_n \rightarrow SU(n)/T(n)$, of
zero-weight bundles over $SU(n)/T(n)$.
 
This point of view is elaborated below.

\subsection{Generalised Weyl group}\label{sec:generalised Weyl}

Let $G$ be a compact, connected semisimple Lie group with maximal
torus $T$. Let $N$ denote the normaliser of $T$, and $W = N/T$ the
Weyl group of $G$.  Suppose $G$ is a subgroup of a compact, connected
Lie group $F$.  Let $M$ denote the $F$-normaliser of $T$, ie the
subgroup of $F$ which leaves $T$ invariant under conjugation.
%\begin{equation}
%  \label{eq:F - normaliser}
%  M = \{x \in F \vert xTx^{-1} = T\}.
%\end{equation}
We call 
\begin{equation}
  \label{eq:generalised Weyl group}
  V = M/T
\end{equation}
the generalised Weyl group of $G$.  Clearly 
the Weyl group $W$ is a subgroup of $V$.

Let $Z$ denote the $F$-centraliser of $T$, ie the subgroup of $F$
whose elements commute with all elements of $T$. $Z$ is a normal
subgroup of $M$.  Therefore, the group $ZN$, consisting of products
$zy$ of $z \in Z$ and $y \in N$, is a subgroup of $M$, and
%$ZN$ is 
%isomorphic to $Z\rtimes N/W$, where the multiplication
%The quotient $(Z N)/T$ is
%therefore a subgroup of $V$.
%It is straightforward to show that
\begin{equation}
  \label{eq: (Z_F N)}
  ZN/ T \cong Z/T \rtimes W,
\end{equation}
where in the semidirect product $Z/T \rtimes W$, an element $y\,T \in W$
acts on $z\,T \in Z/T$ according to $z\,T \rightarrow (yzy^{-1})\,T$.  The
isomorphism \eqref{eq: (Z_F N)} follows from consideration of the map
\begin{equation}
  \label{eq:mapping}
  zy \in  Z N \mapsto (z\,T,y\,T) \in  Z/T \rtimes W.
\end{equation}
To show that this map is well defined, we need to check that $zy =
z'y'$ implies that $z\,T = z'\,T$ and $y\,T =y'\,T$.  But $zy = z'y'$ implies
that $z' = z\tau$ and $y' = \tau^{-1} y$, where $\tau \in Z \cap N$.
Since $G$ is compact and connected, $Z \cap N = T$, so $\tau \in
T$ as required.  It is evident that the map preserves multiplication
and that it is surjective. The kernel of the map consists of elements
$zy$ where $z,y \in T$, and therefore is just $T$ itself, so that  \eqref{eq:
  (Z_F N)} follows.

The spin-statistics group $\Sigma(n)$ is an example of a generalised
Weyl group, with $F = SU(2n)$, $G = SU(n)$, and $T = T(n)$. The Weyl
group $W$ is isomorphic to $S_n$, $Z/T$ is isomorphic to $\Spn(n)$,
and $V$, the generalised Weyl group, is isomorphic to the semidirect
product $\Spn(n) \rtimes S_n$, which is just $\Sigma(n)$.

\subsection{Zero-weight representations of the generalised Weyl group} 
\label{sec:generalised zero weight repn}

Given $T \subset G \subset F$ and $N \subset M$ as above.  Let $\falg$
denote the (real) Lie algebra of $F$, $\Exp:\falg \rightarrow F$
the exponential map, and $\ad$ the adjoint
representation of $F$ on $\falg$.  Let $\galg \subset \falg$ denote
the Lie algebra of $G$, and $\talg \subset \galg$ the Lie algebra of
$T$, (ie, the Cartan subalgebra of $G$), with dual $\talg^*$.  Denote
the pairing between $\mu\in \talg^*$ and $\tau\in \talg$ by $\mu
\cdot\tau$. The adjoint representation 
restricts to a representation of $M$ on $\talg$, denoted
$\ad(M)$.  The co-adjoint representation of $M$ on $\talg^*$, denoted
$\ad^*(M)$, is defined by
\begin{equation}\label{eq:coadj}
(\ad^*(x)\cdot\mu)\cdot \tau  =
\mu \cdot (\ad(x)\cdot\tau).
\end{equation}
Let 
$\ker_\talg(\Exp)$ denote the lattice in $\talg$ mapped to the identity
in $T$.  A weight $\mu$ of $T$ is an element of $\tau^*$ which is
integer-valued on $\ker_\talg(\Exp)$. Irreducible representations of $T$ are
labeled by weights, and are given explicitly by 
$\Exp\tau \mapsto \exp(2\pi \ui \mu\cdot \tau)$.

Let $\Gamma$ denote an irreducible unitary representation of $F$ on a
finite-dimensional Hilbert space $\Vcal$. $\Vcal$ may be decomposed
into a direct sum of generalised weight spaces $\Vcal^\mu$ on which
$\Gamma(T)$ acts with weight $\mu$. (In case $F = G$, this is the
usual weight-space decomposition of $\Vcal$.). Let $\Vcal^0$ denote
the zero-weight space, ie the subspace of vectors invariant under
$T$, for which $\mu = 0$.

For $x \in M$ and $\Exp\tau\in T$, we have that 
\begin{eqnarray}
  \label{eq:action on weights}
  \Gamma(\Exp\tau)\cdot(\Gamma(x)\cdot\Vcal^\mu) &=&
\Gamma(x) \cdot (\Gamma(\Exp (\ad(x^{-1})\cdot \tau)) \cdot \Vcal^\mu) \nonumber \\
&=& \exp(2\pi \ui \mu \cdot (\ad(x^{-1}\cdot \tau)) \Gamma(x)  \cdot \Vcal^\mu
\nonumber \\
&=& \exp(2\pi \ui (\ad^*(x^{-1})\cdot\mu) \cdot t) 
(\Gamma(x)  \cdot \Vcal^\mu),
\end{eqnarray}
so that
\begin{equation}
  \label{eq:weight space transform}
  \Gamma(x) \cdot \Vcal^\mu = \Vcal^{{\rm ad}^*(x^{-1})\cdot \mu}.
\end{equation}
It follows that the zero-weight space, $\Vcal^0$, is invariant under $M$, so
that $\Gamma$ restricts to a representation of $M$ on $\Vcal^0$.
Since $T$ is contained in the kernel, $\Gamma(M)$
reduces to a representation of the generalised Weyl group $V = M/T$ on
$\Vcal^0$.  Denote this representation by $\Delta^\Gamma$.  In general,
$\Delta^\Gamma$ is reducible.  Let $\Delta$ denote an
irreducible representation of $\Vcal$, and let $\nu(\Gamma,\Delta)$
denote the multiplicity of $\Delta$ in the decomposition of
$\Delta^\Gamma$ into its irreducible components. 
The multiplicities $\nu(\Gamma,\Delta)$ are naturally associated with
a pair of irreducible representations $\Gamma$ and $\Delta$ of
a compact connected Lie group $F$ and the generalised Weyl group $V$.
A natural question is how to compute them.  In case $F = G = SU(n)$, this
question has been discussed by Kostant \cite{paper:kostant}.

\subsection{Weight bundles over $G/T$}\label{weight bundles}
% Let $G$ be a Lie group with  unitary representation $\Gamma$ on a

Associated to the weight space $\Vcal^\mu$ is a vector bundle
$\Ecal^\mu$ over $G/T$.  $\Ecal^\mu$ is a sub-bundle of the trivial
bundle $G/T \times \Vcal$, with fibres $\Ecal_{g\, T}$ given by
$\Gamma(g)\cdot \Vcal^\mu$.  By virtue of its embedding in the trivial
bundle, there is an induced $G$-invariant connection on $\Ecal^\mu$,
according to which a vector $\ket{\psi(t)} \in \Ecal_{g(t)\, T}$ 
is parallel transported
along a curve $g(t)\, T\in G/T$ if and only if $\ket{\dot\psi(t)}$
is orthogonal (with respect to the inner product on $\Vcal$) to the
fibre $\Ecal_{g(t)\, T}$.

It is straightforward to derive an explicit formula for parallel
transport along a one-parameter subgroup,
\begin{equation}
  \label{eq:one param}
  g(t) = \Exp(t\xi),
\end{equation}
where
$\xi\in \galg$.  We note that $\Gamma$ gives a representation on $\Vcal$ of
$\falg$, and, by restriction, of $\galg$, by anti-hermitian linear
transformations. We denote these Lie-algebra representations by $\Gamma$
as well. Let $\talgperp \subset \galg$ denote the orthogonal
complement of $\talg$ in $\galg$ with respect to the Killing form (as $G$ is
compact and semisimple, the Killing form is negative definite). It is
a standard result (see eg \cite{book:duistermaatkolk}) that
$\Gamma(\talgperp)$ maps $\Vcal^\mu$ into a direct sum of orthogonal
subspaces $\Vcal^{\mu'}$ (the difference $\mu' - \mu$ is, in fact, a
root of $\galg$). Given $\xi \in \galg$, let $\xi^\talg +
\xi^{\talgperp}$ denote its (unique) decomposition into components
in $\talg$ and $\talgperp$ respectively.  Then, for $\ket{\psi}\in
\Vcal^\mu$, we have that
\begin{equation}
  \label{eq:parallelcomponent}
  \Gamma(\xi)\ket{\psi} = 2\pi \ui (\mu \cdot \xi^\talg)\ket{\psi} + 
\{\hbox{\rm
  vectors orthogonal to $\Vcal^\mu$}\}
\end{equation}
It follows from \eqref{eq:parallelcomponent} that the parallel
transport of $\ket{\psi}$ along $g(t)$
is given by
\begin{equation}
  \label{eq:left action}
  \ket{\psi(t)} = 
\exp(- 2\pi \ui \mu\cdot \xi^\talg) \Gamma(g(t)) \ket{\psi}.
\end{equation}
It follows that the induced connection~\eqref{eq:left action} is Abelian; under
parallel transport around a closed curve in $G/T$, a vector in $\Ecal^\mu$
returns to itself up to a phase factor.  

Let
\begin{eqnarray}
\label{eq:invariant vector fields}
X_\xi(g\,T)& = &\frac{d}{dt}\left.\Exp(t\xi)g\,T\right |_0,\nonumber\cr
X_\eta(g\,T)& = &\frac{d}{dt}\left.\Exp(t\eta)g\,T\right |_0
\end{eqnarray}
denote tangent vector fields at $g\,T$ generated by the left action of $G$.
From \eqref{eq:left action} one can deduce that the scalar-valued
curvature two-form $\Omega^\mu$ on $X_\xi$, $X_\eta$  is given by
\begin{equation}
  \label{eq:curvature}
  \Omega^\mu(X_\xi,X_\eta)(g\,T) = \ui 
\mu\cdot \left([\xi,\eta]^\talg - [\xi^\talg,\eta^\talg]\right).
\end{equation}
Since the left-invariant vector fields span the tangent bundle of $G/T$,
\eqref{eq:curvature} determines $\Omega^\mu$. The curvature form, like the
connection, is invariant under the action of $G$.  

\subsection{Zero-weight bundle and representations of the generalised 
Weyl group}
\label{zero-weight bundles}

Suppose the representation $\Gamma$ of $F$ has a nontrivial
zero-weight space $\Vcal^0$.  From \eqref{eq:curvature}, the curvature
of the associated zero-weight bundle $\Ecal^0$ vanishes, so that
induced connection on $\Ecal^0$ is flat.  In this case, parallel
transport with respect to a flat connection depends only on the
homotopy class of the path in $G/T$.  If $G/T$ is simply connected,
parallel transport is path independent, and $\Ecal^0$ is globally
flat, and therefore trivial.  This is the case if $G$ itself is simply
connected, as we will assume from now on.  As $G$ is compact and
connected, $g(t)\in G$ can be expressed $\Exp (t\xi(t))$ for some
$\xi(t)\in\galg$.  It follows from \eqref{eq:left action} that
parallel transport along in $\Ecal$ along $g(t)\, T$ is given by
\begin{equation}
  \label{eq:zero parallel transport}
  \ket{\phi(t)} = \Gamma(g(t))\ket{\phi}.
\end{equation}

The zero-weight bundle $\Ecal^0$, in contrast to weight bundles with
non-zero weights, descends from a bundle over $G/T$ to a bundle over
$G/N$.  We denote this reduced bundle by $\Ecalbar^0$.  $\Ecalbar^0$
is a sub-bundle of the trivial bundle $G/N\times \Vcal$, with fibres
given by $\Ecalbar_{g\,N} = \Gamma(g)\Vcal^0$.  (Note that since $N$
leaves $\Vcal^0$ invariant, this expression does not depend on the
choice of representative $g$ for $g\, N$.)  The flat connection on
$\Ecal$ passes to $\Ecalbar$.

In general, the $G/N$ is not simply connected; its fundamental group
is isomorphic to the Weyl group $W = N/T$, as follows from the fact 
that $G/N =
(G/T)/(N/T) = (G/T)/W$, and $G/T$ is simply connected by assumption.
An isomorphism between $W$ and
$\pi_1(G/N, N)$, the fundamental group based at the identity coset
$\UI\,N$, is given explicitly as follows.  Let
$g(t)\,N$  denote a closed path in $G/N$ beginning and ending at
$N$. For definiteness, take $0\le t\le 1$ and $g(0) = \UI$.  Then 
$g(1) \,N = \UI \, N$ implies that $g(1)\in N$.  The map
\begin{equation}
  \label{eq:iso fund weyl}
  g(t)\, N\mapsto g(1) \, T
\end{equation}
depends only on the homotopy class of homotopy class of $g(t)\, N$.
It is easily verified that \eqref{eq:iso fund weyl} preserves group
multiplication, and is $1-1$ (since $T$ is connected and $G$ is simply
connected) and onto (since $G$ is connected).

Because $G/N$ is not simply connected, parallel transport with respect
to the flat connection need not be trivial, and can depend on the
homotopy class of the path.  For closed paths, parallel transport
generates a unitary representation of the fundamental group, 
the monodromy of the connection.  In view of the preceding, the monodromy
at the identity coset $N$ is naturally regarded as a representation of
the Weyl group $W$.  We denote this representation by $\Delta^\Gamma$,
and compute it as follows.  Given $y \in N$, let $g(t) \in G$ be a
smooth path in $G$ with $g(0) = \UI$ and $g(1) = y$.
From \eqref{eq:zero parallel transport},
parallel transport in $\Ecalbar$ along $g(t)\, N$ is given by
\begin{equation}
  \label{eq:zero parallel transport quotient}
  \ket{\psi(t)} = \Gamma(g(t))\ket{\psi}.
\end{equation}
$\Delta^\Gamma$ is obtained from parallel transport at $t = 1$, so
that
\begin{equation}
  \label{eq:monodromy appendix}
  \Delta^\Gamma(y) = \Gamma(y).
\end{equation}
This is just the restriction to $N$ of the representation
$\Delta^\Gamma$ of the generalised Weyl group $V$ on $\Ecalbar_N$.

At an arbitrary fibre $\Ecalbar_{g\, N}$ of the quotient bundle we
can define a unitary representation of the generalised
Weyl group $V$, which we denote by
$L_{g\,N}(y)$.  For example, we can take $L_{g\,N}(y) =
\Gamma(g) \Gamma(y) \Gamma^{\dag}(g)$. A different choice of
representative $g$ for $g\,N$ would yield a different but
equivalent representation.  Therefore, there is a well-defined
decomposition of $\Ecalbar$ into a direct sum of sub-bundles
$\Ecalbar^\alpha$ whose fibres transform according to irreducible
representations $\Delta$ of $V$.  This decomposition is
determined by the multiplicities $\nu(\Gamma,\Delta)$ discussed in
Section~\ref{sec:generalised Weyl}.

\bibliography{../../reffs/papers.bib,../../reffs/books.bib}

\end{document}